\def\RELEASE{0}  %
\def\ANON{0}     %
\def\SQUEEZE{0}  %

\documentclass[letterpaper,twocolumn,10pt]{article}
\usepackage{usenix2020-09}
\PassOptionsToPackage{hyphens}{url}\usepackage{hyperref}

\usepackage[noend]{algpseudocode}
\usepackage{algorithmicx}
\usepackage{booktabs}
\usepackage{caption}
\usepackage{comment}
\usepackage[inline]{enumitem}
\usepackage{graphicx}
\usepackage{listings}
\usepackage{multirow}
\usepackage{xcolor}
\usepackage{xspace}
\usepackage[tt=false]{libertine}
\usepackage{inconsolata}
\usepackage{microtype}
\usepackage{tikz}
\usepackage{paralist}
\usepackage{titlesec}
\usepackage{gnuplot-lua-tikz}
\usepackage{xcolor}

\microtypecontext{spacing=nonfrench}

\hypersetup{
  pdflang={en},
  pdfdisplaydoctitle}

\definecolor[named]{OurPurple}{cmyk}{0.55,1,0,0.15}
\definecolor[named]{OurDarkBlue}{cmyk}{1,0.58,0,0.21}
\hypersetup{
  colorlinks,
  linkcolor=OurPurple,
  citecolor=OurPurple,
  urlcolor=OurDarkBlue,
  filecolor=OurDarkBlue}

\if \SQUEEZE 1
\setlist[itemize]{
  leftmargin=*,
  itemsep=2pt,
  topsep=2pt}

\fi

\titlespacing*{\paragraph}{0pt}{2mm}{2mm}
\def\Snospace~{\S{}}

\if \RELEASE 1
  \def\NOTES{0}
\else
  \def\NOTES{1}
\fi
\if \NOTES 1
  \newcommand{\XXX}[1]{{\color{red}{XXX {#1}}}}
  \newcommand{\matheus}[1]{{\color{violet}{[\textbf{MS:} {#1}]}}}
  \newcommand{\antoine}[1]{{\color{teal}{[\textbf{AK:} {#1}]}}}
  \newcommand{\todo}[1]{{\color{blue}{TODO: {#1}}}}
\else
  \newcommand{\XXX}[1]{}
  \netcommand{\matheus}[1]{}
  \newcommand{\antoine}[1]{}
  \newcommand{\todo}[1]{}
\fi

  \newcommand{\sys}{Chamelio\xspace}

\lstdefinelanguage{C}{
  alsoletter={_},
  morekeywords={
    proto_t,auto,break,case,char,const,continue,default,do,double,else,enum,extern,
    float,for,goto,if,inline,int,long,register,restrict,return,short,signed,
    sizeof,static,struct,switch,typedef,union,unsigned,void,volatile,while,
    _Alignas,_Alignof,_Atomic,_Bool,_Complex,_Generic,_Imaginary,_Noreturn,
    _Static_assert,_Thread_local,sentry_t,qentry_t,proto_t,sched_t,
  },
  sensitive=true,
  morecomment=[l]{//},
  morecomment=[s]{/*}{*/},
  morestring=[b]",
}

\definecolor{myblue}{HTML}{4870dd}
\definecolor{mypink}{HTML}{c02d71}
\lstMakeShortInline[columns=fixed]|

\begin{document}
\date{}
  \title{\sys: A Fast Shared Cloud Network Stack for Isolated Tenant-Defined Protocols}
  \author{
    Matheus Stolet\\
    Max Planck Institute for Software Systems
    \and
    Simon Peter\\
    University of Washington
    \and
    Antoine Kaufmann\\
    Max Planck Institute for Software Systems
    }

\maketitle

\begin{abstract}
Conventional cloud network virtualization sends packets through multiple guest
and host layers, inflating CPU cost and tail latency.
Shared host datapaths collapse this layering into one optimized path across
tenants, but existing shared stacks are fixed-function: tenants cannot
specialize their protocols.
eBPF is the natural vehicle for restoring programmability to a shared
datapath, but today’s extensions are hook-sized, and its verifier provides
safety---not performance isolation: one tenant’s per-packet work can inflate
every other tenant’s tail latency.

\sys is a programmable shared network stack that lets tenants implement full
protocols through a bounded eBPF fast path and a tenant slow path,
while approaching the performance and preserving the strong isolation of fixed
shared stacks.
It combines three ideas: a shared-stack architecture for tenant-defined
protocols; joint optimisation of tenant handlers with provider infrastructure
and co-resident tenants in the shared fast path; and a bounded fast path
contract with runtime cycle accounting that keeps tenant programmability
compatible with strong performance isolation.
A tenant-programmable TCP on \sys reaches 9.2\,Mreq/s, matching the
hand-tuned TAS stack; joint compilation shrinks the programmability tax from
23.9\% to 3.8\%; and under a scaling TCP adversary that drives uninstrumented
stacks to 154\,\textmu s, \sys bounds victim tail latency at 46\,\textmu s.
\end{abstract}

\section{Introduction}

Cloud tenants want custom transports---QUIC variants,
RDMA-over-Ethernet, Homa~\cite{montazeri:homa}, domain-specific congestion
control---but today they must choose between two unattractive options.
Provider-managed shared stacks~\cite{marty:snap,dalton:andromeda} consolidate
packet processing onto a small host datapath, where locality, batching, and
cross-tenant core pooling keep per-packet cost low. They are fast and
efficient, but provider-controlled and fixed-function.
Layered hypervisor stacks and per-VM userspace
stacks~\cite{jeong:mtcp,peter:arrakis,zhang:demikernel} restore flexibility but
abandon the locality and pooling that made the shared path efficient.
The missing design point is a shared cloud stack that preserves fast path
efficiency while letting tenants specialise protocols to their applications.

Reaching that point is hard because flexibility tends to destroy what makes
shared fast paths fast.
Once protocol logic is exposed through hooks or layered boundaries, the
datapath is no longer compiled as one streamlined path: extensions incur
dispatch costs and are opaque to end-to-end optimisation.
Where common-case packet work already runs in hundreds of cycles, those
overheads erode both throughput and latency.

Sharing a programmable fast path also creates a performance-isolation
problem: tenant code on shared cores lets one tenant's per-packet work
inflate every other tenant's tail latency.
eBPF is the natural vehicle for untrusted tenant code in networking
datapaths~\cite{jorgensen:xdp}, but its verifier targets safety, not
performance isolation---instruction bounds are a coarse proxy for cycle
cost and ignore helper overheads and cross-tenant contention.
Runtime cycle accounting has isolated shared fast paths before, but only
for fixed provider-defined code (Virtuoso~\cite{stolet:virtuoso}); extending
it to tenant eBPF, whose cost varies widely within the safety envelope,
requires co-designing the programming model and the bounded execution
contract needed for isolation.

\sys is a programmable shared network stack for cloud VMs that exposes a
provider-managed shared host datapath as a substrate for tenant-defined
protocols.
Existing programmable stacks either run per VM
(Demikernel~\cite{zhang:demikernel}, mTCP~\cite{jeong:mtcp}) or expose only
hook-sized extensions (XDP~\cite{jorgensen:xdp}), while existing
shared stacks remain provider-controlled and fixed-function
(Snap~\cite{marty:snap}, Virtuoso~\cite{stolet:virtuoso}).
Two mechanisms make tenant-programmable sharing viable in \sys.
First, a control plane jointly compiles tenant eBPF handlers with provider
infrastructure code---and across co-resident tenants---recovering the
end-to-end optimisation that isolated JIT and extension dispatch would
otherwise destroy.
Second, \sys constrains the fast path contract to three bounded
common-case handlers for receive, scheduler-triggered send, and dequeue
events (RX, SCHED, DEQ) that execute close to the NIC over
protocol-local state in per-VM shared memory, while complex or state-heavy
logic stays in a tenant-resident slow path. In other words, \sys does not
implement an entire protocol in eBPF: only the latency-critical common path
runs in bounded host handlers.
Runtime cycle accounting over this bounded contract makes tenant-defined
eBPF compatible with strong shared-stack performance isolation.

Our evaluation shows \sys delivers three properties usually in tension.
First, \sys is expressive enough for full protocols at state-of-the-art
performance: \sys UDP reaches 3.7\,Mreq/s ($7.9\times$
Demikernel~\cite{zhang:demikernel}), and \sys TCP reaches
9.2\,Mreq/s---matching the hand-tuned TAS~\cite{kaufmann:tas} stack while
remaining tenant-programmable.
Second, cross-layer joint compilation cuts the programmability tax from
23.9\% to 3.8\% for TCP.
Third, bounded handlers plus runtime accounting preserve isolation: under
a scaling TCP adversary, an uninstrumented victim's tail latency rises
from 14\,\textmu s to 154\,\textmu s, but \sys bounds it at
46\,\textmu s---and our sensitivity study shows even complex protocols
like TCP fit within the bounded fast path regime needed for this isolation.

We make the following contributions:
\begin{compactitem}[\labelitemi]
\item A \emph{shared-stack architecture for tenant protocols} that exposes
a provider-managed host datapath as a programmable substrate for full
protocols in virtualised environments.
\item \emph{Cross-layer, cross-tenant fast path optimisation} that jointly
compiles tenant eBPF handlers with provider infrastructure and co-resident
tenant handlers into a single streamlined datapath.
\item \emph{Performance isolation for tenant-defined protocols on a shared
fast path}, using a bounded handler contract and runtime cycle accounting
to make tenant eBPF compatible with strong shared-stack isolation.
\end{compactitem}

\noindent
This work complies with all ethical standards of the home institutions.
\sys will be open-sourced on publication.
\section{Background and Motivation}%
\label{sec:bg}

In this section, we motivate the use of shared datapaths
in virtualised cloud environments because of their resource efficiency
and explain how they cause performance interference between tenants
and limit protocol flexibility. Next, we show how
the extended Berkeley Packet Filter (eBPF) enables network
datapath programmability but does not yet offer
an interface that allows tenants to build full
protocols. Finally, we present the challenges to
building shared network datapaths for VMs that enforce
isolation between tenants, support full tenant-defined protocols,
and achieve performance on par with optimised fixed fast paths.

\subsection{Shared Network Datapaths}

\begin{figure}%
\centering
\includegraphics[width=0.46\textwidth]{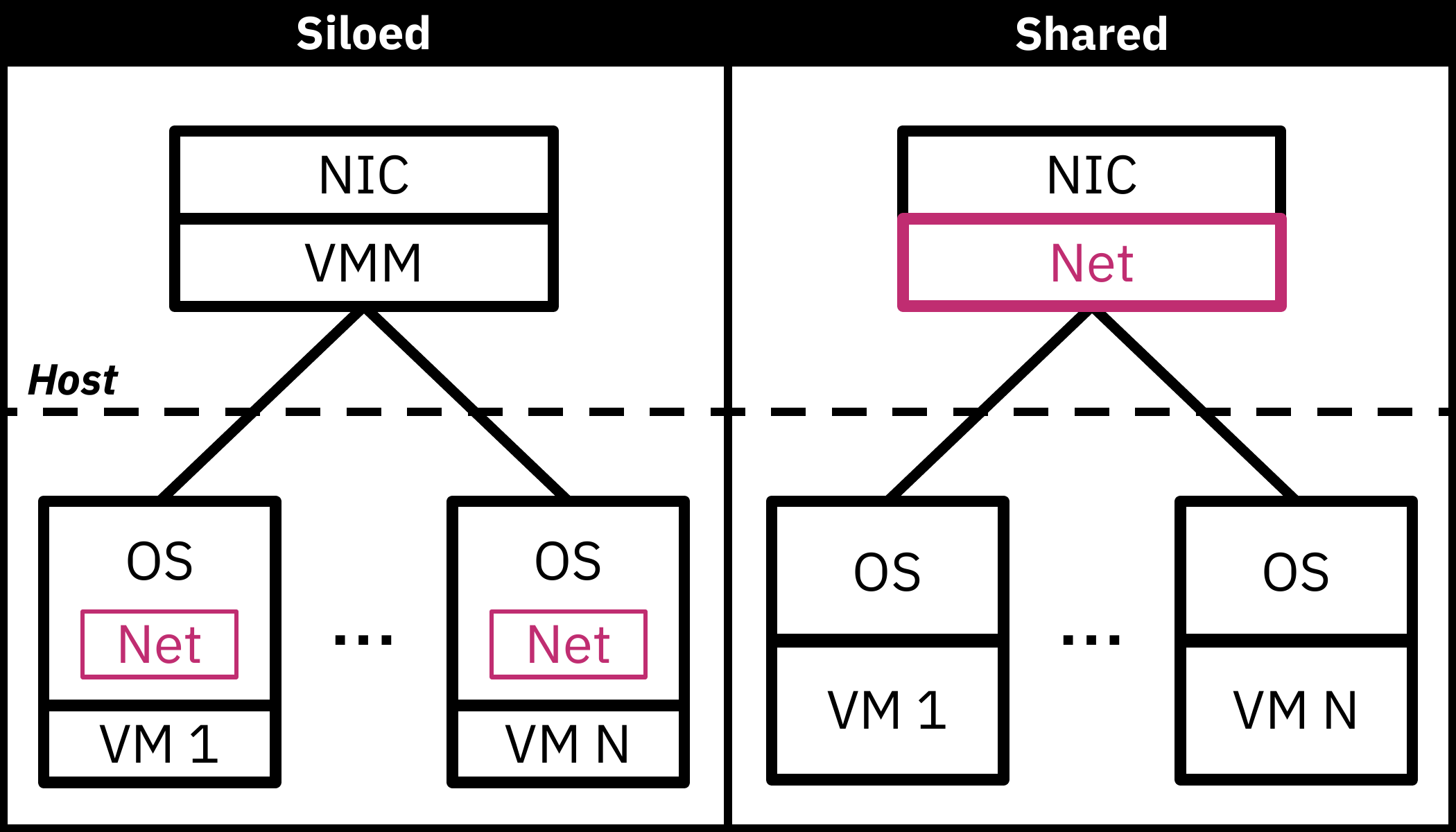}%
\caption{Per-guest datapaths are the norm, but shared datapaths across guests
promise greater efficiency.}%
\label{fig:shared_stack}%
\end{figure}

Shared network datapaths, illustrated in \autoref{fig:shared_stack},
consolidate VM networking into a
single shared instance, offering substantial efficiency benefits
in virtualised environments. This consolidation enables more
effective use of CPU resources through smoother handling of bursty workloads.
However, shared datapaths also introduce challenges arising from multi-tenancy.

\textbf{Shared network datapaths improve efficiency
in virtualised environments.}
Shared network datapaths have been deployed
in production~\cite{dalton:andromeda,marty:snap} because they
allow multiple guests and tenants to share a
single datapath instance. This consolidation improves
CPU utilisation by elastically allocating resources
rather than statically provisioning per-VM peak capacity.
Shared datapaths also enable aggressive batching, which amortises
fixed per-packet overheads across bursts and keeps hot code
and metadata resident in cache.
Prior work has shown that the
capacity to elastically allocate
resources and absorb bursts can improve CPU resource efficiency
by 82\% when compared to siloed per-VM datapaths
and can save over 40\% CPU cores for the operator without
performance degradation when multiplexing VMs~\cite{niu:netkernel}.

\textbf{Sharing complicates performance isolation.}
Despite resource efficiency benefits, shared datapaths must fairly 
allocate CPU and network resources
across mutually untrusted guests~\cite{stolet:virtuoso,dalton:andromeda}.
Multiple tenants share a core, so excess CPU consumption by one tenant
directly delays service to others. For example, a datapath may service
multiple VM queues in the same thread, so a burst from one VM occupies
this thread and delays the service of other VMs.

\textbf{Untrusted tenants constrain protocol flexibility.}
In addition to complicating performance isolation,
shared datapaths limit protocol flexibility.
Datacentre applications are constantly evolving and becoming
increasingly diverse. Developers thus write custom
protocols to improve
performance and add custom features to meet changing
demands.
The issue is that, in cloud environments where tenants are untrusted,
operators cannot safely execute tenant-defined protocols in a
shared datapath running on the host.
In contrast, with siloed datapaths, tenants can safely
modify the guest OS or run userspace network stacks
inside the VM to implement tenant-defined protocols; the VM isolation
boundaries prevent one tenant from interfering with
another.

In shared kernel datapaths, where such isolation is absent,
developers rely on eBPF to introduce extensibility.
However, as we show in the next section,
this approach provides limited support for
performance isolation between tenants, incurs
additional execution overheads,
and lacks sufficient expressivity to support full
protocols.

\subsection{eBPF in Network Datapaths}

The extended Berkeley Packet Filter (eBPF) is used to enable
programmable packet processing in Linux networking.
It lets operators inject small, verified, JIT-compiled programs
at well-defined points in the datapath to implement custom filtering,
steering, encapsulation, and observability. This subsection explains
how eBPF is used in today's network stack,
the programming and execution model that makes it safe and fast,
and finally why this form of extensibility
by itself does not provide full protocols with
strong performance isolation in shared datapaths.

\textbf{eBPF makes datapaths safely programmable.}
The technology behind eBPF enables users to extend and specialise
the Linux networking datapath to their
applications without forking the kernel
or rewriting the stack. For example, XDP~\cite{jorgensen:xdp}
builds on eBPF to create an early hook that moves
user-defined logic closer to the NIC and
supports processing ahead of the main networking stack.
Systems have leveraged this insight to specialise the
network datapath to individual applications with low overhead.
For example, Electrode~\cite{zhou:electrode}
offloads distributed protocols into eBPF to reduce
kernel crossings and BMC~\cite{ghigoff:bmc}
boosts Memcached performance by serving hot keys from
an in-kernel cache.
To understand how this extensibility is safely done in practice
and the restrictions it imposes,
we next look at the eBPF programming model and toolchain.

\textbf{Programs are written in restricted C and compiled to bytecode.}
Userspace programs use the LLVM eBPF backend to compile C programs
into bytecode that contains native eBPF CPU instructions that execute
in the kernel environment.
The C code interacts with the host environment through
a narrow interface that uses helpers and maps, rather than arbitrary
pointer chasing. This interface imposes restrictions that limit
the expressivity and hamper the complexity of the resulting program,
but ensures its safety and facilitates automatic verification.

\textbf{Static verifier enforces safety.}
Instead of relying on an interpreter, eBPF enforces safety
with a static verifier so that it can execute native CPU
instructions and improve performance~\cite{jacobson:bsd_packet_filter}.
The verifier checks that the loaded
bytecode can only read and write to memory that the program owns.
Additionally, the verifier ensures that the bytecode
terminates within a certain number of instructions, guaranteeing
that the kernel is not starved of resources.
If the bytecode passes the verifier checks, the kernel
JITs and loads the resulting eBPF program into memory,
where it attaches the program to hook points.

\textbf{Linux runs eBPF at networking hook points.}
After the eBPF program is loaded into kernel memory,
the userspace program specifies an event and the kernel attaches
the program to that event. For example, the execution
of a system call or the arrival of a packet at the NIC are events that
can trigger an eBPF program. Driver hooks,
such as XDP, move processing before a packet reaches the
network stack; socket hooks enable custom policies per
connection; Linux traffic control hooks permit custom
traffic shaping. In short, eBPF enables extensibility
of the Linux datapath, but this extensibility is
restricted to small logic that runs in these hook points.

\textbf{Protocols outgrow hook-sized programs.}
The Linux datapath already supports eBPF, but its
extensibility is not designed for full protocols
and strong performance isolation. For instance, complex
protocols want richer composition, state, and control flow than
typical in-kernel eBPF usage. Furthermore, verifier safety
does not ensure performance isolation caused by
contention of shared resources; shared datapaths must then
support a mix of tenant-defined protocols without
causing performance interference.

\textbf{Userspace eBPF broadens extensions.}
Tools such as rBPF~\cite{software:rbpf}, uBPF~\cite{software:ubpf},
and bpftime~\cite{zheng:bpftime} use eBPF bytecode to
run verifiable extensions inside userspace applications
and services. They preserve a similar toolchain and programming
model as kernel eBPF, but allow developers to attach logic at
function and library boundaries instead of kernel events.
Userspace eBPF creates opportunities to extend applications
beyond kernel hookpoints, but current
tools do not eliminate the trade-offs between extensibility, isolation,
and performance. Existing systems such as rBPF and uBPF provide
incomplete eBPF support and higher overhead, while bpftime
relies on additional runtime mechanisms to enforce safety
and isolation in userspace.

\subsection{Challenges for Shared Datapaths in Virtualised Environments}

Building a shared network datapath for virtualised
environments raises three
core challenges. This datapath must offer a programming model
that allows tenants to implement full protocols
instead of small extensions. It must also enforce strict
security and performance isolation for untrusted protocol
logic, while achieving performance that is competitive with highly
optimised fixed fast paths.

\paragraph{C1: Safely exposing full-protocol programmability.}
Complex protocols are stateful and
span many events, so they do not fit neatly
into tiny per-packet
eBPF hooks~\cite{shahinfar:ebpf_app_perf,miano:complex_services}.
These hook-based extensions fragment protocol logic across
callbacks, obscuring control flow, limiting composition,
and breaking the compiler's view of the datapath.
A model that is too restrictive limits protocol complexity.
This demands an interface that can express full protocols,
such as TCP and UDP, yet keeps the verified and trusted surface small.

\paragraph{C2: Enforcing performance isolation with
  tenant-provided implementations.}
The eBPF verifier enforces the safety of tenant-supplied protocol
code, but safe code can still act as a noisy neighbour
by consuming excessive CPU cycles or triggering behaviour
that degrades the performance of other
tenants~\cite{shahinfar:ebpf_app_perf}. Existing mechanisms, such as
loop termination checks and instruction limits,
do not provide sufficient isolation. Programs invoke helper
functions whose costs are not accounted for during verification,
and worst-case execution time~\cite{sahu:bpf_mgmt,wilhelm:wcet} estimates derived
from control-flow analysis remain imprecise.
Admission enforcement is thus not enough to guarantee performance isolation
and needs to be coupled with runtime enforcement.

\paragraph{C3: Achieving performance on par with optimised fixed
fast paths.}
eBPF can introduce overheads such as dispatch,
validation, and boundary
crossings~\cite{shahinfar:ebpf_app_perf,liu:ebpf_maps,joly:tail_call_costs,miano:nitrosketch}.
These overheads quickly
dominate per-packet cost.
Furthermore, eBPF programs are compiled independently from
the underlying network stack; this separation makes the
eBPF program opaque to the compiler and forgoes optimisation
opportunities across boundaries in the eBPF program and
fixed network datapath that are typically applied to conventionally compiled code.
Achieving fixed-datapath performance while still enabling
tenant-specific behaviour is therefore essential.
These challenges directly shape \sys's design: the next section shows how
\sys uses a bounded host-side contract, joint compilation, and runtime cycle
accounting to support full tenant-defined protocols with strong
isolation while retaining shared-stack performance.
\section{\sys Design}%
\label{sec:design}

\sys targets a missing design point in cloud networking:
it should preserve the efficiency of a provider-managed shared host datapath
while allowing tenants to deploy full custom protocols.
Achieving this requires balancing multiple goals that are usually in tension.
This section first states these goals and then explains how \sys meets them
through a cross-trust-boundary architecture, a protocol decomposition that
places bounded common-case logic on the host and complex logic in the guest,
joint compilation, and runtime cycle accounting.

\subsection{Design Goals}
\label{sec:design_goals}

\begin{compactitem}[\labelitemi]
\item \textbf{High shared-stack performance.}
The common-case datapath should retain the locality, batching, and
cross-tenant core-pooling benefits of provider-managed shared stacks.

\item \textbf{Full-protocol flexibility.}
Tenants should be able to implement complete protocols,
not just hook-sized extensions at fixed points in the datapath.

\item \textbf{Strong isolation for untrusted tenants.}
Tenant protocol logic must preserve safety and avoid unbounded
performance interference with co-resident tenants.

\item \textbf{Independent evolution across the trust boundary.}
Operators should be able to evolve shared infrastructure independently,
while tenants can deploy and update protocol logic without modifying the
provider stack.

\item \textbf{A narrow trusted host interface.}
The programmable surface exposed on the host should stay small and bounded
enough to verify, optimise, and schedule safely.
\end{compactitem}

These goals shape the rest of the design.
At a high level, \sys defines where the system components run, how protocols
are partitioned across them, how tenant fast path code is admitted and
optimised, and how runtime CPU consumption is accounted for to bound
interference.

\subsection{Architectural Overview}
\label{sec:design_architecture}

\sys comprises three architectural components:
a provider-managed fast path on the host,
a tenant-resident slow path in the guest,
and a control plane that admits, verifies, and compiles tenant fast path code.
These components execute in separate protection domains and interact
through a narrow interface based on bounded handlers and per-VM shared state.
This design exposes the shared host datapath as a programmable substrate
while preserving isolation and efficiency across the tenant/provider boundary.

  \begin{figure*}%
  \centering
  \includegraphics[width=0.8\textwidth]{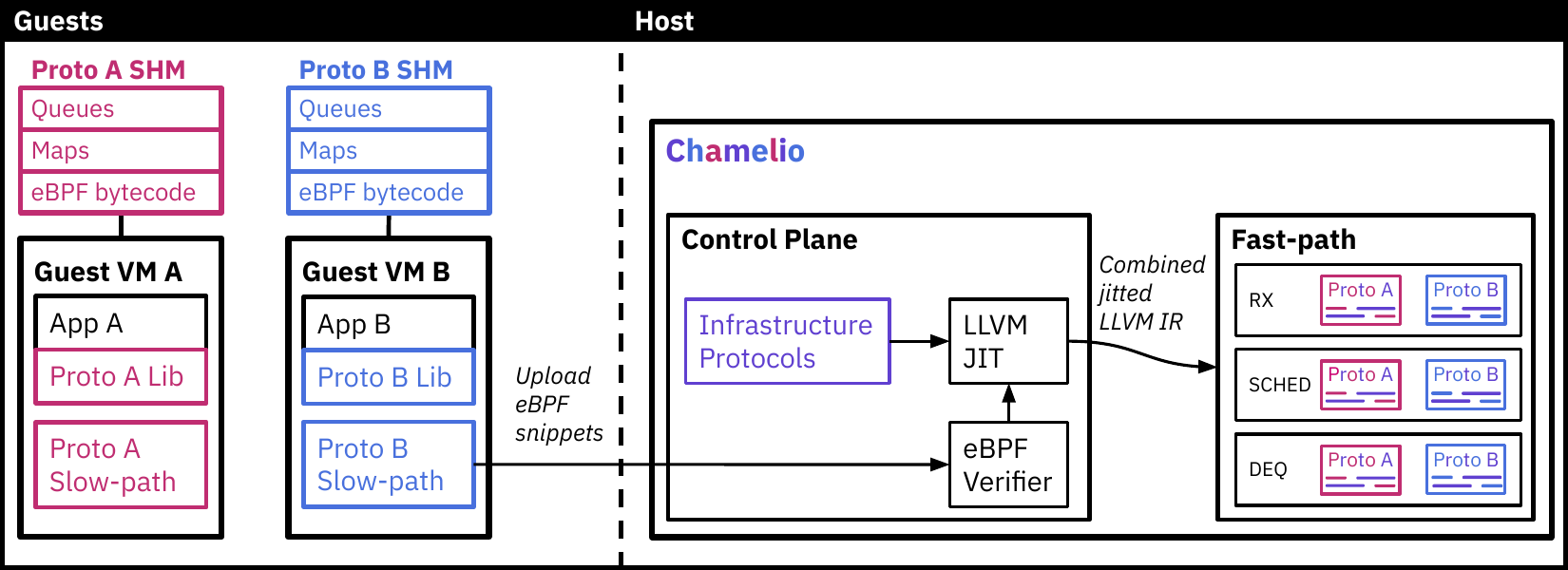}%
  \caption{\sys architecture. Each guest gets an exclusive shared-memory region for protocol
  state and handler upload. The control plane verifies uploaded eBPF and
  installs specialised RX, SCHED, and DEQ fast path entry points.}%
  \label{fig:architecture}%
  \end{figure*}

\paragraph{Host fast path.}
The host fast path executes tenant-defined protocol handlers
for different guests in a shared datapath. It runs as a polling
loop close to the NIC, batching packets across tenants for
better efficiency. This design preserves the key efficiency
benefits of shared stacks: it maintains locality of code
and data, enables batching across tenants, and amortises
per-packet overheads. By keeping common-case packet processing
on the host and close to the NIC, \sys avoids the layering
of per-VM datapaths, while still allowing tenants to customise
protocol behaviour through bounded handlers.

\paragraph{Tenant slow path.}
The tenant slow path runs inside the guest VM and handles complex,
infrequent, or state-heavy protocol operations that would bloat the
latency-critical fast path. By keeping these operations in the guest,
\sys maintains a small and verifiable host interface while allowing
tenants to implement full protocols. This separation preserves the
efficiency of the shared datapath and avoids exposing complex logic to
the trusted host environment. As a result, performance-critical
handlers execute on the host, while richer protocol functionality
remains in the guest, where tenants retain full flexibility beyond
the constraints of the eBPF verifier.

\paragraph{Control plane.}
The control plane is the host-side component that manages tenant
fast path code over time, separate from both the packet-processing
fast path and the tenant-resident protocol logic. Guests use it to
upload and replace bounded handlers, and the control plane installs the
resulting specialised entry points in the fast path. The control plane also
allocates each VM's shared-memory region and maintains the host-side
state needed to manage active tenant protocols. Verification,
compilation, and hot updates are handled by this component but are
described in detail in the later subsections.

\paragraph{Per-VM shared state.}
\sys assigns each guest an exclusive shared memory region that
serves as the boundary between the host datapath and tenant
protocols. The control plane allocates this region and exposes it
inside the VM as a PCIe device, allowing the guest to map it directly
while preserving isolation across tenants.
Both fast path handlers and the tenant slow path access this region to
store protocol state, exchange messages, and pass queue entries.
This design enables low-overhead
coordination between host and guest while keeping the interface narrow
and well-defined.
By keeping protocol state in exclusive per-VM shared-memory regions,
\sys preserves VM isolation and lets tenants evolve their protocols
independently without modifying the shared host datapath.

\subsection{Protocol Decomposition and Fast Path Contract}
\label{sec:design_decomposition}

The central design question is how to split a tenant protocol across
the host and guest while preserving both performance and safety.
While prior systems separate common-case fast path work from rare or
complex slow path logic, \sys performs this decomposition across the
tenant/provider trust boundary and must therefore expose a host-side
contract that is both narrow and expressive enough
for full protocols. \sys addresses this challenge with three bounded
eBPF handlers—\texttt{event\_rx}, \texttt{event\_sched}, and
\texttt{event\_deq}—that capture common-case protocol processing
for packet reception, scheduler-triggered send, and dequeue
events.

\paragraph{Why these three handlers.}
\texttt{event\_rx}, \texttt{event\_sched}, and \texttt{event\_deq}
capture the latency-critical common case across network protocols
because they align with the three recurring
events that drive packet processing: packet arrival, the scheduler
releasing staged work, and the delivery of work from the application or slow
path. This contract is a better fit for full protocols than
conventional hook-sized extensions because it groups related work into
a small number of semantically meaningful phases, rather than
fragmenting protocol logic across many narrow callbacks tied to
individual kernel hook points. As a result, tenants can implement full
protocols with explicit receive, scheduler-triggered send, and dequeue paths,
while the
host still exposes a narrow and bounded interface for the
common case.
Conventional eBPF interfaces expose narrow, event-specific hook points.
This works for small packet-processing extensions, but it fragments
protocols that must coordinate state across receive,
scheduler-triggered send, and control events.

\paragraph{What stays in the fast path.}
The fast path executes bounded eBPF handlers that operate on the
common-case per-packet processing path. These handlers perform
packet parsing and classification, construct and modify protocol
headers, make queueing and scheduling decisions, and update
protocol state stored in per-VM shared memory. For example,
the \texttt{event\_rx} handler parses incoming packets and maps
them to protocol state, \texttt{event\_sched} constructs outgoing
packets using state retrieved from shared memory, and
\texttt{event\_deq} processes messages from the slow path and
applications. Many protocols also need one event to trigger future handler
executions, for example to pace a later segment transmission or to break one
logical operation into multiple follow-on events. Expressing such chains
directly inside bounded eBPF would require loops or long-lived control flow
that do not fit the fast path model. \sys therefore provides a small
host-side scheduler, manipulated through helpers, that stages work across the
RX, SCHED, and DEQ phases without inflating a single handler. In particular, work
staged by one handler can later be popped by the scheduler, which triggers
\texttt{event\_sched} to materialise the next packet. All operations are
designed to complete within a bounded number of instructions, ensuring
predictable execution while keeping packet processing close to the NIC. This
restriction keeps the fast path streamlined and defers complex or state-heavy
logic to the tenant slow path.

\paragraph{What stays in the slow path.}
The slow path handles protocol logic that falls outside the bounded,
per-packet execution model of the fast path. This includes connection
setup and teardown, retransmission timeouts, and other state-heavy
operations that require complex control
flow or long-lived state. These operations execute inside the guest VM,
where they can access and update protocol state stored in the per-VM
shared memory region and communicate with the fast path via queues.
Because they are infrequent and not on the critical packet-processing
path, they do not need to satisfy the strict execution bounds imposed
on fast path handlers. Keeping this logic in the guest avoids bloating
the trusted host interface while allowing tenants to implement full
protocols with rich state and control flow.

\paragraph{Why this is sufficient for full protocols.}
UDP and TCP illustrate why the three handlers are sufficient for full
protocols. In both cases, the common path fits within the bounded host-side
contract, while more complex or infrequent logic remains in the guest slow
path. In UDP,
the common case is message-oriented: \texttt{event\_rx} parses and
demultiplexes incoming packets, while \texttt{event\_deq} consumes
application or slow path messages and can emit packets directly without
staging through \texttt{event\_sched}. In TCP, the common case also
decomposes into the same bounded phases. \texttt{event\_rx} processes
incoming segments and updates per-connection state in shared memory.
\texttt{event\_deq} handles bumps and control messages and stages future
transmissions in the scheduler. When that staged work is later popped,
\texttt{event\_sched} constructs the corresponding data packets and ACKs.
Operations that do not fit this bounded
contract, such as connection setup and teardown, retransmission
timeouts, and other rare or state-heavy cases, remain in the tenant
slow path. As a result, the full protocol spans both host and guest,
while its latency-critical transport logic fits within a small,
verifiable host-side interface.

\subsection{Fast Path Code Admission and Static Safety}
\label{sec:design_admission}

Tenants deploy host-side protocol logic by uploading eBPF bytecode for
the bounded handlers, but this code executes on shared cores in the
trusted datapath. The control plane therefore enforces safety
and bounded execution before any code is installed
in the fast path. It verifies memory safety and termination, enforces
static limits on execution cost, and constructs the combined program
that will execute in the datapath, ensuring that all tenant-provided
code conforms to a narrow, well-defined contract prior to execution.

\paragraph{Upload and registration.}
The guest slow path uploads bounded handler bytecode to the control plane,
which registers it in the fast path after verification. The slow
path first requests shared-memory space for its eBPF bytecode and then
writes the handler programs into that region. It then sends the control
plane a message requesting installation. When a guest uploads new eBPF
code, the control plane recompiles the aggregate module, including
infrastructure code and all registered programs, and then updates the
fast path's handler entry functions. Decoupling the control
plane from the fast path allows us to do this without
affecting packet processing that is already running.

\paragraph{Memory safety and termination.}
When a slow path uploads custom
eBPF programs for |event_rx|, |event_sched|,
and |event_deq|,
the control plane guarantees the
safety of these programs by verifying them with the
Prevail~\cite{gershuni:prevail} user-space verifier.
The verifier builds a control-flow graph by converting bytecode into
structured instructions and validates each instruction for
memory bounds, arithmetic safety, and size constraints.
For load/store instructions that access shared memory,
the verifier checks each access against memory region bounds,
preventing one protocol from reading or writing another VM's memory.
The verifier also checks the control-flow graph for loops to ensure
proper program termination.

\paragraph{Static bounds.}
The verifier imposes a maximum number of instructions per fast path
program.
It analyses the bytecode, generates a control flow graph, and then
computes the longest execution path for an uploaded program.
The control plane rejects programs that
exceed an operator-defined maximum.
This maximum protects guests from performance interference
by preventing long-running programs from monopolizing
resources.
However, instruction count is a coarse proxy for execution time and does not
capture the full cost of machine code and helper calls, as instructions
vary in cost. To address this, we complement static bounds with
fine-grained, per-protocol cycle-level accounting and runtime
enforcement in the fast path, providing stronger isolation in the datapath.

\subsection{Joint Compilation}
\label{sec:design_compilation}

Programmability should not force tenant protocol logic to run as an opaque
callback separated from provider infrastructure by dispatch and abstraction
boundaries.
\sys therefore links tenant handlers, active guest protocols, and shared
fast path infrastructure into one LLVM module, optimises across
tenant/provider and tenant/tenant boundaries in that combined program, and
emits specialised entry functions for each fast path phase. This is the
mechanism behind the low programmability tax measured in
\autoref{ssec:eval:joint_compilation}.

\paragraph{Unified IR module.}
After verification, we convert the newly uploaded bytecode to LLVM IR
instead of immediately lowering it to machine code. To do this, we extend
\textit{llvmbpf}~\cite{zheng:bpftime} to emit LLVM IR for the uploaded program.
The control plane then links that IR with precompiled infrastructure modules
(e.g. ARP handling and network virtualisation), helper implementations, and
the IR of active guest protocols from co-resident tenants in the same shared
fast path. The result is one aggregate module representing the full fast path
for that shared-stack instance rather than a set of opaque callbacks.

\paragraph{Cross-boundary optimisation.}
Before machine code generation, we internalise infrastructure, helper, and
eBPF definitions across the aggregate module. This makes those functions
local to the module, allowing the compiler to inline across tenant/provider
and tenant/tenant boundaries and eliminate now-redundant definitions. We then
run a host-aware, post-link interprocedural optimisation pipeline over the
full call graph. Together, these passes collapse callback layers, propagate
constants, and avoid unnecessary stores to shared state and reloads across the combined fast path.

\paragraph{Specialised stage entry points.}
Code generation emits one entry function per fast path phase. Instead of a
generic polling loop dispatching into opaque per-guest handlers, the fast path
executes specialised RX, SCHED, and DEQ stage functions over the combined code
path. For example, when the scheduler-triggered send stage computes context
metadata and
the handler immediately consumes it, the compiler can keep that value in a
register instead of storing it in the context and reloading it. The resulting
machine code is therefore much closer to a hand-specialised implementation than
to an isolated extension callback, directly reducing the residual
programmability tax.

\subsection{Runtime Isolation}
\label{sec:design_isolation}

Static verification alone cannot guarantee strong performance isolation,
because machine-code costs vary across instructions and helper calls, and the
shared fast path introduces cross-tenant contention at runtime.
\sys therefore complements admission-time bounds with runtime cycle
accounting.

\paragraph{Why static bounds are insufficient.}
Static verification ensures that programs are safe and terminate,
but it does not capture their execution cost accurately.
Instruction bounds are a coarse proxy for runtime: different
instructions and helper calls incur different costs, and
microarchitectural contention changes those costs at runtime.
Furthermore, static analysis cannot account for
cross-tenant contention: even bounded handlers can inflate the tail
latency of other tenants when they execute on shared cores.
Therefore, admission-time bounds must be complemented with
runtime enforcement that measures and controls per-protocol resource
consumption.

\paragraph{Cycle accounting and budgets.}
\sys enforces isolation by allocating each guest a token budget,
where each token represents one TSC cycle in the fast path.
The fast path measures cycles spent executing each protocol's
handlers by reading the TSC before and after execution and charges the
elapsed time to that protocol's budget.
By default, guests receive an equal share of CPU cycles, although allocations can be weighted
to reflect operator policy. Budgets are capped to prevent protocols from
accumulating credit while idle and then bursting, and are refilled
periodically by the control plane every 1,\textmu s.

\paragraph{Scheduling under contention.}
If a protocol exhausts its budget, the fast path skips it in
subsequent poll rounds until the control plane refills its credits and
the fast path continues servicing other eligible protocols. This
mechanism bounds the work each tenant can perform in the datapath and
prevents any single protocol from causing unbounded interference. In
addition, the instruction limits enforced by the verifier keep each
handler short enough to run to completion without preemption. This
avoids preemption overheads while still supporting strong isolation.

\subsection{Programming Interface}
\label{sec:design_interface}

\sys exposes distinct control plane and fast path
interfaces. Protocol authors see three abstractions: shared maps and queues in
per-VM memory, three bounded fast path handlers, and a small helper set.
The control plane interface lets the slow path create protocol state and upload
handlers, while the fast path interface defines what host-resident
handlers can access on each event. Protocol code does not manipulate arbitrary
host memory or NIC internals directly; it operates on packet buffers,
bounds-checked shared memory, dequeued work items, and helper-mediated state
access.

\paragraph{Control plane interface.}
The control plane interface in \autoref{fig:control_interface} is used by the
slow path to create a protocol instance, allocate shared protocol state, and upload
fast path bytecode. |cham_new_proto| maps the per-VM shared-memory region into
the guest process. |cham_map| and |cham_queue| allocate protocol state and
communication channels in that region. |cham_enable_queue| and
|cham_disable_queue| control whether the fast path polls a queue and thus
whether dequeues from it can trigger |event_deq|. |cham_alloc_ebpf|,
|cham_up_ebpf|, and |cham_free_ebpf| manage the uploaded handler bytecode.

\begin{figure}[!t]
\centering
\begin{lstlisting}

// Creates proto and maps shared memory
proto_t cham_new_proto(guest);

// Creates a queue in shared memory
int cham_queue(proto, nelems, elsize);
// Creates a map in shared memory
int cham_map(proto, nelems, elsize);

// Enables queue in fast path core
int cham_enable_queue(proto, qid, core);
// Disables queue from fast path
int cham_disable_queue(proto, qid);

// Allocates space for eBPF bytecode
int cham_alloc_ebpf(proto, size);
// Writes eBPF bytecode to shared memory
int cham_up_ebpf(proto, bcode, size);
// Frees space used by eBPF bytecode
int cham_free_ebpf(proto);

\end{lstlisting}
\caption{Control plane interface.}
\label{fig:control_interface}
\end{figure}

\paragraph{Fast path interface.}
The fast path interface in \autoref{fig:fast_interface} consists of the fixed
call signatures of the tenant's eBPF handlers and the |cham_ctx| passed to
them. Protocol authors implement up to three handlers: |event_rx|,
|event_sched|, and |event_deq|. |event_rx| runs on packet
arrival, |event_sched| when the scheduler pops staged work to send, and
|event_deq| when the slow path or application hands work to the fast path.
Each handler returns the length of a packet to send, or zero if it emits no
packet. When the return value is non-zero, the fast path sends |ctx->pkt| with
that length after the handler returns. |cham_ctx| exposes only the packet
buffer and its bounds, the protocol's shared-memory region and its bounds, a
dequeued queue entry, and scheduler state. In |event_rx|, |ctx->pkt| points to the
received packet. In |event_sched|, |ctx->pkt| points to the packet buffer being
filled. In |event_deq|, |ctx->qe| points to the dequeued work item, and |ctx->pkt| can
be used for direct packet generation. |ctx->sched| is the per-protocol
scheduler state whose queued work triggers |event_sched|. This
keeps the host-side surface narrow while still giving handlers the state they
need for common-case transport processing.

\begin{figure}[!t]
\centering
\begin{lstlisting}
// Context passed to fast path programs
struct cham_ctx {
  // Pointer to packet memory buffer
  void *pkt;
  // End of memory buffer
  void *pkt_end;
  // Pointer to protocol shared memory
  void *shm_base;
  // Shared memory length
  void *shm_end;
  // Dequeued queue entry
  qentry_t *qe;
  // Staged scheduler work
  sched_t sched;
}

// Process received packet
int event_rx(cham_ctx);
// Process scheduler-triggered send work
int event_sched(cham_ctx);
// Process dequeued message
int event_deq(cham_ctx);
\end{lstlisting}
\caption{Fast path interface.}
\label{fig:fast_interface}
\end{figure}

\paragraph{Helpers.}
\sys provides the small helper surface in \autoref{fig:helpers} as the set of
host-provided functions that tenant eBPF handlers may call. Queue and map
helpers perform bounded accesses without exposing internal representations,
while scheduler helpers manipulate the host-side scheduler that stages future
work across the RX, SCHED, and DEQ phases and later triggers |event_sched| by
popping staged work. In |cham_ctx|, |ctx->sched| is the
handle to this per-protocol scheduler state. This surface is enough to express
the protocols we evaluate: UDP uses |event_rx| to demultiplex packets and
|event_deq| to emit replies directly, while TCP uses |event_rx| to update
per-connection state, |event_deq| to stage sends and control messages, and
|event_sched| to construct data packets and ACKs.

\begin{figure}[!t]
\centering
\begin{lstlisting}
// Adds element to queue
int queue_enq(queue, qentry);
// Pops element from queue
qentry_t queue_deq(queue, elsz);
// Looks up an element in a map
int map_lookup(map_base, id, elsz);
// Pops entry from scheduler
sentry_t sched_pop(sched);
// Adds entry to scheduler
int sched_add(sched, id, priority, val);
\end{lstlisting}
\caption{Helper functions.}
\label{fig:helpers}
\end{figure}
\section{Implementation}%
\label{sec:impl}
We implemented \sys from scratch in 7,008 lines of C
code (LoC). The implementation includes 2,082 LoC in the fast path,
3,248 LoC in the control plane. An extra 2,245 LoC were used in the
programming libraries. We also implemented full UDP and TCP on top of
\sys's bounded host fast path and guest slow path. UDP requires 2,313
LoC, including 301 for eBPF handlers, 766 for the guest slow path,
and 933 for the application library.
TCP is larger, at a total of 6,314 LoC,
with 1,046 for eBPF handlers,
2,930 for the guest slow path, and 1,631 for the library.

\sys runs as a userspace process with separate threads for the
fast path and control plane.
The fast path runs as a dedicated polling thread, while
the control plane executes asynchronously to avoid interfering
with packet processing.
We use DPDK~\cite{software:dpdk} to directly access the NIC
and avoid kernel overheads such as interrupts
and system calls.
For tenant-provided host-side handlers, \sys uses a fully userspace eBPF
toolchain built from the Prevail~\cite{gershuni:prevail} verifier and
\textit{llvmbpf}~\cite{zheng:bpftime}. We extend \textit{llvmbpf} to emit LLVM
IR instead of immediately lowering bytecode to machine code, enabling the
joint-compilation pipeline described in \autoref{sec:design_compilation}.

\textbf{Static Instruction Bound Estimation.}
Prevail reports only the largest upper bound on loop iterations.
We extend it to compute an upper bound on the number of
instructions executed along the worst-case path, which we use for the
admission-time static bound.
Our implementation operates over Prevail's control-flow graph.
For acyclic regions, we compute the longest path to program exit by
counting the number of instructions on each path.
For cyclic regions, we conservatively multiply the loop bound by the
total number of instructions in the loop body. This over-approximates
execution cost, as it counts instructions on mutually exclusive paths.
A tighter bound would require path-sensitive analysis across iterations,
tracking how control flow evolves with program state. We leave a more
complex estimation for future work.

\textbf{VM integration.}
\sys integrates with the hypervisor to map shared memory regions
into guest VMs via a dummy PCI device. Inside the VM, the protocol
slow path discovers this device and exposes it as a Unix socket
and shared memory file descriptors. This abstraction avoids
requiring applications to distinguish between native and VM
execution. No modifications are required to the guest OS.
We implement this integration using QEMU~\cite{software:qemu}
with KVM and its Inter-VM Shared Memory (IVSHM) device~\cite{software:ivshm}.
IVSHM allows an external process to pass
a shared-memory file descriptor, which is exposed in the VM as
a PCI device with a directly mapped BAR region. The protocol
slow path maps this memory using \sys libraries, which rely on
\texttt{vfio-pci}~\cite{software:vfio-pci} to implement a
userspace PCI driver. \texttt{vfio-pci} provides a file
descriptor that is \texttt{mmap}'ed for direct access to the
shared memory region. This design enables a shared-memory region
between \sys on the host and applications in the guest,
keeping host/guest coordination to low-overhead shared-memory
accesses.
\section{Evaluation}%
\label{sec:eval}
We implement full UDP and DCTCP-based TCP protocols on \sys and compare them
against optimised baselines. Our evaluation addresses the following questions:

\begin{compactitem}[\labelitemi]
  \item Can \sys support full message and stream transports at
  competitive end-to-end performance? (\autoref{ssec:eval:fullprotos})
  \item How much does joint compilation reduce the programmability tax of
  tenant-defined fast path handlers? (\autoref{ssec:eval:joint_compilation})
  \item Does \sys maintain shared-stack isolation with an
  adversary running tenant-defined handlers?
  (\autoref{ssec:eval:isolation})
  \item Are static admission bounds necessary for isolation, and do real
  protocols fit within the required bound? (\autoref{ssec:eval:static_bounds})
\end{compactitem}

\subsection{Experimental Setup}

\paragraph{Testbed.}
We use two servers connected back-to-back via 
100 Gbps Mellanox ConnectX-5 NICs. Each server is 
a dual-socket Intel Xeon Gold 6152 machine 
(2.1\,GHz, 22 cores per socket; 187\,GB RAM). 
Both run Debian 11 with Linux 6.1.

\paragraph{Baselines.}
We compare against: (1) Linux UDP/TCP; 
(2) Demikernel UDP~\cite{zhang:demikernel}; and 
(3) TAS TCP~\cite{kaufmann:tas}.
Linux, Demikernel, and TAS baselines run inside QEMU/KVM VMs with
SR-IOV~\cite{spec:intel_sriov} virtual functions,
which provide near bare-metal performance by allowing
direct NIC access from the guest.
In \sys, the control plane and fast path run on the host,
while applications and the slow path execute inside the
guest VM; applications interact with \sys via QEMU's
IVSHM~\cite{software:ivshm}. For TCP experiments, 
we use DCTCP as the congestion
control algorithm. TAS and \sys use a single datapath core,
while applications scale across cores to increase load
and saturate the network stack. Since Demikernel
only supports single-threaded applications, we restrict
UDP experiments to one application core for fairness.

\paragraph{Application.}
RPCs are a common and demanding workload
in datacentre applications~\cite{seemakhupt:rpc-charac},
as they are both latency- and throughput-sensitive. We evaluate performance
using an echo server that replies to small 
64-byte messages under increasing request rates, 
stressing the packet-processing path at high packet rates.

\subsection{Full Tenant-Defined Protocols at Shared-Stack Performance}
\label{ssec:eval:fullprotos}

We evaluate whether \sys's bounded receive, scheduler-triggered send, and
dequeue fast path contract (RX, SCHED, DEQ) plus guest slow path support full
tenant-defined protocols without giving up the efficiency of a shared
host stack. TCP is the harder stream-based case; UDP is the complementary
message-based case.

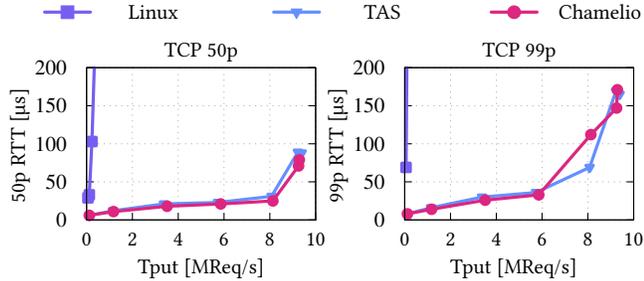
\begin{figure}%
	\centering%
\begin{tikzpicture}[gnuplot]
\tikzset{every node/.append style={font={\fontsize{8.0pt}{9.6pt}\selectfont}}}
\path (0.000,0.000) rectangle (8.458,3.378);
\gpcolor{color=gp lt color border}
\gpcolor{rgb color={0.471,0.369,0.941}}
\gpsetlinewidth{1.00}
\gpsetpointsize{6.00}
\gp3point{gp mark 5}{}{(0.422,3.169)}
\gpcolor{color=gp lt color border}
\gpcolor{rgb color={0.392,0.561,1.000}}
\gp3point{gp mark 11}{}{(3.467,3.169)}
\gpcolor{color=gp lt color border}
\gpcolor{rgb color={0.863,0.149,0.498}}
\gp3point{gp mark 7}{}{(6.359,3.169)}
\gpcolor{color=gp lt color border}
\node[gp node center] at (1.564,3.169) {Linux};
\node[gp node center] at (4.609,3.169) {TAS};
\node[gp node center] at (7.501,3.169) {\sys};
\gpcolor{rgb color={0.471,0.369,0.941}}
\gpsetlinetype{gp lt border}
\gpsetdashtype{gp dt solid}
\gpsetlinewidth{3.00}
\draw[gp path](0.118,3.170)--(0.727,3.170);
\gpcolor{rgb color={0.392,0.561,1.000}}
\draw[gp path](3.163,3.170)--(3.772,3.170);
\gpcolor{rgb color={0.863,0.149,0.498}}
\draw[gp path](6.055,3.170)--(6.664,3.170);
\gpdefrectangularnode{gp plot 1}{\pgfpoint{0.422cm}{2.905cm}}{\pgfpoint{8.034cm}{3.377cm}}
\gpcolor{color=gp lt color axes}
\gpsetlinetype{gp lt axes}
\gpsetdashtype{gp dt axes}
\gpsetlinewidth{0.50}
\draw[gp path] (0.676,0.405)--(3.721,0.405);
\gpcolor{color=gp lt color border}
\gpsetlinetype{gp lt border}
\gpsetdashtype{gp dt solid}
\gpsetlinewidth{1.00}
\draw[gp path] (0.676,0.405)--(0.766,0.405);
\node[gp node right] at (0.529,0.405) {$0$};
\gpcolor{color=gp lt color axes}
\gpsetlinetype{gp lt axes}
\gpsetdashtype{gp dt axes}
\gpsetlinewidth{0.50}
\draw[gp path] (0.676,0.912)--(3.721,0.912);
\gpcolor{color=gp lt color border}
\gpsetlinetype{gp lt border}
\gpsetdashtype{gp dt solid}
\gpsetlinewidth{1.00}
\draw[gp path] (0.676,0.912)--(0.766,0.912);
\node[gp node right] at (0.529,0.912) {$50$};
\gpcolor{color=gp lt color axes}
\gpsetlinetype{gp lt axes}
\gpsetdashtype{gp dt axes}
\gpsetlinewidth{0.50}
\draw[gp path] (0.676,1.418)--(3.721,1.418);
\gpcolor{color=gp lt color border}
\gpsetlinetype{gp lt border}
\gpsetdashtype{gp dt solid}
\gpsetlinewidth{1.00}
\draw[gp path] (0.676,1.418)--(0.766,1.418);
\node[gp node right] at (0.529,1.418) {$100$};
\gpcolor{color=gp lt color axes}
\gpsetlinetype{gp lt axes}
\gpsetdashtype{gp dt axes}
\gpsetlinewidth{0.50}
\draw[gp path] (0.676,1.925)--(3.721,1.925);
\gpcolor{color=gp lt color border}
\gpsetlinetype{gp lt border}
\gpsetdashtype{gp dt solid}
\gpsetlinewidth{1.00}
\draw[gp path] (0.676,1.925)--(0.766,1.925);
\node[gp node right] at (0.529,1.925) {$150$};
\gpcolor{color=gp lt color axes}
\gpsetlinetype{gp lt axes}
\gpsetdashtype{gp dt axes}
\gpsetlinewidth{0.50}
\draw[gp path] (0.676,2.431)--(3.721,2.431);
\gpcolor{color=gp lt color border}
\gpsetlinetype{gp lt border}
\gpsetdashtype{gp dt solid}
\gpsetlinewidth{1.00}
\draw[gp path] (0.676,2.431)--(0.766,2.431);
\node[gp node right] at (0.529,2.431) {$200$};
\gpcolor{color=gp lt color axes}
\gpsetlinetype{gp lt axes}
\gpsetdashtype{gp dt axes}
\gpsetlinewidth{0.50}
\draw[gp path] (0.676,0.405)--(0.676,2.431);
\gpcolor{color=gp lt color border}
\gpsetlinetype{gp lt border}
\gpsetdashtype{gp dt solid}
\gpsetlinewidth{1.00}
\draw[gp path] (0.676,0.405)--(0.676,0.495);
\node[gp node center] at (0.676,0.159) {0};
\gpcolor{color=gp lt color axes}
\gpsetlinetype{gp lt axes}
\gpsetdashtype{gp dt axes}
\gpsetlinewidth{0.50}
\draw[gp path] (1.285,0.405)--(1.285,2.431);
\gpcolor{color=gp lt color border}
\gpsetlinetype{gp lt border}
\gpsetdashtype{gp dt solid}
\gpsetlinewidth{1.00}
\draw[gp path] (1.285,0.405)--(1.285,0.495);
\node[gp node center] at (1.285,0.159) {2};
\gpcolor{color=gp lt color axes}
\gpsetlinetype{gp lt axes}
\gpsetdashtype{gp dt axes}
\gpsetlinewidth{0.50}
\draw[gp path] (1.894,0.405)--(1.894,2.431);
\gpcolor{color=gp lt color border}
\gpsetlinetype{gp lt border}
\gpsetdashtype{gp dt solid}
\gpsetlinewidth{1.00}
\draw[gp path] (1.894,0.405)--(1.894,0.495);
\node[gp node center] at (1.894,0.159) {4};
\gpcolor{color=gp lt color axes}
\gpsetlinetype{gp lt axes}
\gpsetdashtype{gp dt axes}
\gpsetlinewidth{0.50}
\draw[gp path] (2.503,0.405)--(2.503,2.431);
\gpcolor{color=gp lt color border}
\gpsetlinetype{gp lt border}
\gpsetdashtype{gp dt solid}
\gpsetlinewidth{1.00}
\draw[gp path] (2.503,0.405)--(2.503,0.495);
\node[gp node center] at (2.503,0.159) {6};
\gpcolor{color=gp lt color axes}
\gpsetlinetype{gp lt axes}
\gpsetdashtype{gp dt axes}
\gpsetlinewidth{0.50}
\draw[gp path] (3.112,0.405)--(3.112,2.431);
\gpcolor{color=gp lt color border}
\gpsetlinetype{gp lt border}
\gpsetdashtype{gp dt solid}
\gpsetlinewidth{1.00}
\draw[gp path] (3.112,0.405)--(3.112,0.495);
\node[gp node center] at (3.112,0.159) {8};
\gpcolor{color=gp lt color axes}
\gpsetlinetype{gp lt axes}
\gpsetdashtype{gp dt axes}
\gpsetlinewidth{0.50}
\draw[gp path] (3.721,0.405)--(3.721,2.431);
\gpcolor{color=gp lt color border}
\gpsetlinetype{gp lt border}
\gpsetdashtype{gp dt solid}
\gpsetlinewidth{1.00}
\draw[gp path] (3.721,0.405)--(3.721,0.495);
\node[gp node center] at (3.721,0.159) {10};
\draw[gp path] (3.721,0.405)--(3.631,0.405);
\draw[gp path] (3.721,0.912)--(3.631,0.912);
\draw[gp path] (3.721,1.418)--(3.631,1.418);
\draw[gp path] (3.721,1.925)--(3.631,1.925);
\draw[gp path] (3.721,2.431)--(3.631,2.431);
\draw[gp path] (0.676,2.431)--(0.676,0.405)--(3.721,0.405)--(3.721,2.431)--cycle;
\gpcolor{rgb color={0.471,0.369,0.941}}
\gpsetlinewidth{3.00}
\draw[gp path] (0.694,0.699)--(0.712,0.739)--(0.747,1.448)--(0.778,2.431);
\gpsetpointsize{4.00}
\gp3point{gp mark 5}{}{(0.694,0.699)}
\gp3point{gp mark 5}{}{(0.712,0.739)}
\gp3point{gp mark 5}{}{(0.747,1.448)}
\gpcolor{rgb color={0.392,0.561,1.000}}
\draw[gp path] (0.712,0.466)--(1.026,0.527)--(1.704,0.618)--(2.422,0.638)--(3.134,0.719)%
  --(3.485,1.307)--(3.535,1.296);
\gp3point{gp mark 11}{}{(0.712,0.466)}
\gp3point{gp mark 11}{}{(1.026,0.527)}
\gp3point{gp mark 11}{}{(1.704,0.618)}
\gp3point{gp mark 11}{}{(2.422,0.638)}
\gp3point{gp mark 11}{}{(3.134,0.719)}
\gp3point{gp mark 11}{}{(3.485,1.307)}
\gp3point{gp mark 11}{}{(3.535,1.296)}
\gpcolor{rgb color={0.863,0.149,0.498}}
\draw[gp path] (0.712,0.466)--(1.033,0.516)--(1.746,0.587)--(2.459,0.618)--(3.152,0.658)%
  --(3.494,1.124)--(3.504,1.205);
\gp3point{gp mark 7}{}{(0.712,0.466)}
\gp3point{gp mark 7}{}{(1.033,0.516)}
\gp3point{gp mark 7}{}{(1.746,0.587)}
\gp3point{gp mark 7}{}{(2.459,0.618)}
\gp3point{gp mark 7}{}{(3.152,0.658)}
\gp3point{gp mark 7}{}{(3.494,1.124)}
\gp3point{gp mark 7}{}{(3.504,1.205)}
\gpcolor{color=gp lt color border}
\gpsetlinewidth{1.00}
\draw[gp path] (0.676,2.431)--(0.676,0.405)--(3.721,0.405)--(3.721,2.431)--cycle;
\node[gp node center,rotate=-270.0] at (-0.218,1.418) {50p RTT [µs]};
\node[gp node center] at (2.198,-0.209) {Tput [MReq/s]};
\node[gp node center] at (2.198,2.677) {TCP 50p};
\gpdefrectangularnode{gp plot 2}{\pgfpoint{0.676cm}{0.405cm}}{\pgfpoint{3.721cm}{2.431cm}}
\gpcolor{color=gp lt color axes}
\gpsetlinetype{gp lt axes}
\gpsetdashtype{gp dt axes}
\gpsetlinewidth{0.50}
\draw[gp path] (4.905,0.405)--(7.949,0.405);
\gpcolor{color=gp lt color border}
\gpsetlinetype{gp lt border}
\gpsetdashtype{gp dt solid}
\gpsetlinewidth{1.00}
\draw[gp path] (4.905,0.405)--(4.995,0.405);
\node[gp node right] at (4.758,0.405) {$0$};
\gpcolor{color=gp lt color axes}
\gpsetlinetype{gp lt axes}
\gpsetdashtype{gp dt axes}
\gpsetlinewidth{0.50}
\draw[gp path] (4.905,0.912)--(7.949,0.912);
\gpcolor{color=gp lt color border}
\gpsetlinetype{gp lt border}
\gpsetdashtype{gp dt solid}
\gpsetlinewidth{1.00}
\draw[gp path] (4.905,0.912)--(4.995,0.912);
\node[gp node right] at (4.758,0.912) {$50$};
\gpcolor{color=gp lt color axes}
\gpsetlinetype{gp lt axes}
\gpsetdashtype{gp dt axes}
\gpsetlinewidth{0.50}
\draw[gp path] (4.905,1.418)--(7.949,1.418);
\gpcolor{color=gp lt color border}
\gpsetlinetype{gp lt border}
\gpsetdashtype{gp dt solid}
\gpsetlinewidth{1.00}
\draw[gp path] (4.905,1.418)--(4.995,1.418);
\node[gp node right] at (4.758,1.418) {$100$};
\gpcolor{color=gp lt color axes}
\gpsetlinetype{gp lt axes}
\gpsetdashtype{gp dt axes}
\gpsetlinewidth{0.50}
\draw[gp path] (4.905,1.925)--(7.949,1.925);
\gpcolor{color=gp lt color border}
\gpsetlinetype{gp lt border}
\gpsetdashtype{gp dt solid}
\gpsetlinewidth{1.00}
\draw[gp path] (4.905,1.925)--(4.995,1.925);
\node[gp node right] at (4.758,1.925) {$150$};
\gpcolor{color=gp lt color axes}
\gpsetlinetype{gp lt axes}
\gpsetdashtype{gp dt axes}
\gpsetlinewidth{0.50}
\draw[gp path] (4.905,2.431)--(7.949,2.431);
\gpcolor{color=gp lt color border}
\gpsetlinetype{gp lt border}
\gpsetdashtype{gp dt solid}
\gpsetlinewidth{1.00}
\draw[gp path] (4.905,2.431)--(4.995,2.431);
\node[gp node right] at (4.758,2.431) {$200$};
\gpcolor{color=gp lt color axes}
\gpsetlinetype{gp lt axes}
\gpsetdashtype{gp dt axes}
\gpsetlinewidth{0.50}
\draw[gp path] (4.905,0.405)--(4.905,2.431);
\gpcolor{color=gp lt color border}
\gpsetlinetype{gp lt border}
\gpsetdashtype{gp dt solid}
\gpsetlinewidth{1.00}
\draw[gp path] (4.905,0.405)--(4.905,0.495);
\node[gp node center] at (4.905,0.159) {0};
\gpcolor{color=gp lt color axes}
\gpsetlinetype{gp lt axes}
\gpsetdashtype{gp dt axes}
\gpsetlinewidth{0.50}
\draw[gp path] (5.514,0.405)--(5.514,2.431);
\gpcolor{color=gp lt color border}
\gpsetlinetype{gp lt border}
\gpsetdashtype{gp dt solid}
\gpsetlinewidth{1.00}
\draw[gp path] (5.514,0.405)--(5.514,0.495);
\node[gp node center] at (5.514,0.159) {2};
\gpcolor{color=gp lt color axes}
\gpsetlinetype{gp lt axes}
\gpsetdashtype{gp dt axes}
\gpsetlinewidth{0.50}
\draw[gp path] (6.123,0.405)--(6.123,2.431);
\gpcolor{color=gp lt color border}
\gpsetlinetype{gp lt border}
\gpsetdashtype{gp dt solid}
\gpsetlinewidth{1.00}
\draw[gp path] (6.123,0.405)--(6.123,0.495);
\node[gp node center] at (6.123,0.159) {4};
\gpcolor{color=gp lt color axes}
\gpsetlinetype{gp lt axes}
\gpsetdashtype{gp dt axes}
\gpsetlinewidth{0.50}
\draw[gp path] (6.731,0.405)--(6.731,2.431);
\gpcolor{color=gp lt color border}
\gpsetlinetype{gp lt border}
\gpsetdashtype{gp dt solid}
\gpsetlinewidth{1.00}
\draw[gp path] (6.731,0.405)--(6.731,0.495);
\node[gp node center] at (6.731,0.159) {6};
\gpcolor{color=gp lt color axes}
\gpsetlinetype{gp lt axes}
\gpsetdashtype{gp dt axes}
\gpsetlinewidth{0.50}
\draw[gp path] (7.340,0.405)--(7.340,2.431);
\gpcolor{color=gp lt color border}
\gpsetlinetype{gp lt border}
\gpsetdashtype{gp dt solid}
\gpsetlinewidth{1.00}
\draw[gp path] (7.340,0.405)--(7.340,0.495);
\node[gp node center] at (7.340,0.159) {8};
\gpcolor{color=gp lt color axes}
\gpsetlinetype{gp lt axes}
\gpsetdashtype{gp dt axes}
\gpsetlinewidth{0.50}
\draw[gp path] (7.949,0.405)--(7.949,2.431);
\gpcolor{color=gp lt color border}
\gpsetlinetype{gp lt border}
\gpsetdashtype{gp dt solid}
\gpsetlinewidth{1.00}
\draw[gp path] (7.949,0.405)--(7.949,0.495);
\node[gp node center] at (7.949,0.159) {10};
\draw[gp path] (7.949,0.405)--(7.859,0.405);
\draw[gp path] (7.949,0.912)--(7.859,0.912);
\draw[gp path] (7.949,1.418)--(7.859,1.418);
\draw[gp path] (7.949,1.925)--(7.859,1.925);
\draw[gp path] (7.949,2.431)--(7.859,2.431);
\draw[gp path] (4.905,2.431)--(4.905,0.405)--(7.949,0.405)--(7.949,2.431)--cycle;
\gpcolor{rgb color={0.471,0.369,0.941}}
\gpsetlinewidth{3.00}
\draw[gp path] (4.923,1.104)--(4.934,2.431);
\gp3point{gp mark 5}{}{(4.923,1.104)}
\gpcolor{rgb color={0.392,0.561,1.000}}
\draw[gp path] (4.941,0.486)--(5.255,0.567)--(5.933,0.709)--(6.650,0.770)--(7.363,1.104)%
  --(7.713,2.117)--(7.763,2.076);
\gp3point{gp mark 11}{}{(4.941,0.486)}
\gp3point{gp mark 11}{}{(5.255,0.567)}
\gp3point{gp mark 11}{}{(5.933,0.709)}
\gp3point{gp mark 11}{}{(6.650,0.770)}
\gp3point{gp mark 11}{}{(7.363,1.104)}
\gp3point{gp mark 11}{}{(7.713,2.117)}
\gp3point{gp mark 11}{}{(7.763,2.076)}
\gpcolor{rgb color={0.863,0.149,0.498}}
\draw[gp path] (4.941,0.486)--(5.262,0.547)--(5.975,0.668)--(6.687,0.739)--(7.380,1.540)%
  --(7.722,1.894)--(7.732,2.137);
\gp3point{gp mark 7}{}{(4.941,0.486)}
\gp3point{gp mark 7}{}{(5.262,0.547)}
\gp3point{gp mark 7}{}{(5.975,0.668)}
\gp3point{gp mark 7}{}{(6.687,0.739)}
\gp3point{gp mark 7}{}{(7.380,1.540)}
\gp3point{gp mark 7}{}{(7.722,1.894)}
\gp3point{gp mark 7}{}{(7.732,2.137)}
\gpcolor{color=gp lt color border}
\gpsetlinewidth{1.00}
\draw[gp path] (4.905,2.431)--(4.905,0.405)--(7.949,0.405)--(7.949,2.431)--cycle;
\node[gp node center,rotate=-270.0] at (4.010,1.418) {99p RTT [µs]};
\node[gp node center] at (6.354,-0.209) {Tput [MReq/s]};
\node[gp node center] at (6.427,2.677) {TCP 99p};
\gpdefrectangularnode{gp plot 3}{\pgfpoint{4.905cm}{0.405cm}}{\pgfpoint{7.949cm}{2.431cm}}
\end{tikzpicture}
\caption{TCP latency and throughput under load; \sys matches TAS at
  9.2\,Mreq/s.}%
\label{fig:tcp_lat_tp}%
\end{figure}

\textbf{TCP.}
TCP is the harder case: it combines connection state, congestion control, and
inter-packet dependencies, so success here is evidence that \sys supports full
stream transports rather than hook-sized extensions. In
\autoref{fig:tcp_lat_tp} we report throughput and latency for \sys, TAS, and
Linux TCP stacks. We omit the Demikernel because it does not support
multi-threaded applications, which are required to saturate TAS.

The \sys TCP implementation matches TAS in throughput,
with both systems
saturating at 9.2 million requests per second. Under
light load, \sys achieves lower latency: at 3.5 million
requests per second, it attains a 50th-percentile RTT of
18\,\textmu s and a 99th-percentile RTT of
26\,\textmu s, compared to 21\,\textmu s and 30\,\textmu s for TAS.
This improvement stems in part from \sys's disaggregated
architecture, where the network stack runs on the host
close to the NIC, while applications execute in the VM.
As a result, certain operations, such as ACKs without
payload, can be handled entirely on the host without
involving the guest.
However, as load increases, \sys exhibits a higher 99th-percentile RTT during
the transition to saturation. At 8 million requests per second, it reaches
112\,\textmu s, compared to 69\,\textmu s for TAS.

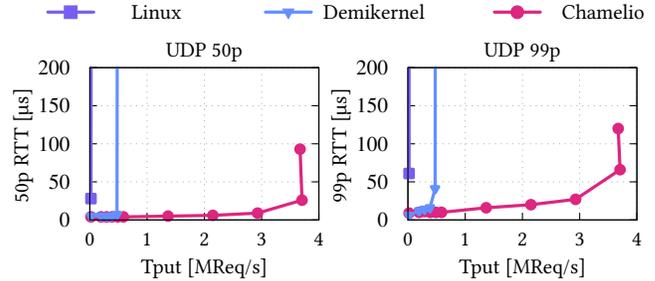
\begin{figure}%
\centering%
\begin{tikzpicture}[gnuplot]
\tikzset{every node/.append style={font={\fontsize{8.0pt}{9.6pt}\selectfont}}}
\path (0.000,0.000) rectangle (8.458,3.378);
\gpcolor{color=gp lt color border}
\gpcolor{rgb color={0.471,0.369,0.941}}
\gpsetlinewidth{1.00}
\gpsetpointsize{6.00}
\gp3point{gp mark 5}{}{(0.422,3.169)}
\gpcolor{color=gp lt color border}
\gpcolor{rgb color={0.392,0.561,1.000}}
\gp3point{gp mark 11}{}{(3.315,3.169)}
\gpcolor{color=gp lt color border}
\gpcolor{rgb color={0.863,0.149,0.498}}
\gp3point{gp mark 7}{}{(6.359,3.169)}
\gpcolor{color=gp lt color border}
\node[gp node center] at (1.564,3.169) {Linux};
\node[gp node center] at (4.456,3.169) {Demikernel};
\node[gp node center] at (7.501,3.169) {\sys};
\gpcolor{rgb color={0.471,0.369,0.941}}
\gpsetlinetype{gp lt border}
\gpsetdashtype{gp dt solid}
\gpsetlinewidth{3.00}
\draw[gp path](0.118,3.170)--(0.727,3.170);
\gpcolor{rgb color={0.392,0.561,1.000}}
\draw[gp path](3.011,3.170)--(3.620,3.170);
\gpcolor{rgb color={0.863,0.149,0.498}}
\draw[gp path](6.055,3.170)--(6.664,3.170);
\gpdefrectangularnode{gp plot 1}{\pgfpoint{0.422cm}{2.905cm}}{\pgfpoint{8.034cm}{3.377cm}}
\gpcolor{color=gp lt color axes}
\gpsetlinetype{gp lt axes}
\gpsetdashtype{gp dt axes}
\gpsetlinewidth{0.50}
\draw[gp path] (0.676,0.405)--(3.721,0.405);
\gpcolor{color=gp lt color border}
\gpsetlinetype{gp lt border}
\gpsetdashtype{gp dt solid}
\gpsetlinewidth{1.00}
\draw[gp path] (0.676,0.405)--(0.766,0.405);
\node[gp node right] at (0.529,0.405) {$0$};
\gpcolor{color=gp lt color axes}
\gpsetlinetype{gp lt axes}
\gpsetdashtype{gp dt axes}
\gpsetlinewidth{0.50}
\draw[gp path] (0.676,0.912)--(3.721,0.912);
\gpcolor{color=gp lt color border}
\gpsetlinetype{gp lt border}
\gpsetdashtype{gp dt solid}
\gpsetlinewidth{1.00}
\draw[gp path] (0.676,0.912)--(0.766,0.912);
\node[gp node right] at (0.529,0.912) {$50$};
\gpcolor{color=gp lt color axes}
\gpsetlinetype{gp lt axes}
\gpsetdashtype{gp dt axes}
\gpsetlinewidth{0.50}
\draw[gp path] (0.676,1.418)--(3.721,1.418);
\gpcolor{color=gp lt color border}
\gpsetlinetype{gp lt border}
\gpsetdashtype{gp dt solid}
\gpsetlinewidth{1.00}
\draw[gp path] (0.676,1.418)--(0.766,1.418);
\node[gp node right] at (0.529,1.418) {$100$};
\gpcolor{color=gp lt color axes}
\gpsetlinetype{gp lt axes}
\gpsetdashtype{gp dt axes}
\gpsetlinewidth{0.50}
\draw[gp path] (0.676,1.925)--(3.721,1.925);
\gpcolor{color=gp lt color border}
\gpsetlinetype{gp lt border}
\gpsetdashtype{gp dt solid}
\gpsetlinewidth{1.00}
\draw[gp path] (0.676,1.925)--(0.766,1.925);
\node[gp node right] at (0.529,1.925) {$150$};
\gpcolor{color=gp lt color axes}
\gpsetlinetype{gp lt axes}
\gpsetdashtype{gp dt axes}
\gpsetlinewidth{0.50}
\draw[gp path] (0.676,2.431)--(3.721,2.431);
\gpcolor{color=gp lt color border}
\gpsetlinetype{gp lt border}
\gpsetdashtype{gp dt solid}
\gpsetlinewidth{1.00}
\draw[gp path] (0.676,2.431)--(0.766,2.431);
\node[gp node right] at (0.529,2.431) {$200$};
\gpcolor{color=gp lt color axes}
\gpsetlinetype{gp lt axes}
\gpsetdashtype{gp dt axes}
\gpsetlinewidth{0.50}
\draw[gp path] (0.676,0.405)--(0.676,2.431);
\gpcolor{color=gp lt color border}
\gpsetlinetype{gp lt border}
\gpsetdashtype{gp dt solid}
\gpsetlinewidth{1.00}
\draw[gp path] (0.676,0.405)--(0.676,0.495);
\node[gp node center] at (0.676,0.159) {0};
\gpcolor{color=gp lt color axes}
\gpsetlinetype{gp lt axes}
\gpsetdashtype{gp dt axes}
\gpsetlinewidth{0.50}
\draw[gp path] (1.437,0.405)--(1.437,2.431);
\gpcolor{color=gp lt color border}
\gpsetlinetype{gp lt border}
\gpsetdashtype{gp dt solid}
\gpsetlinewidth{1.00}
\draw[gp path] (1.437,0.405)--(1.437,0.495);
\node[gp node center] at (1.437,0.159) {1};
\gpcolor{color=gp lt color axes}
\gpsetlinetype{gp lt axes}
\gpsetdashtype{gp dt axes}
\gpsetlinewidth{0.50}
\draw[gp path] (2.199,0.405)--(2.199,2.431);
\gpcolor{color=gp lt color border}
\gpsetlinetype{gp lt border}
\gpsetdashtype{gp dt solid}
\gpsetlinewidth{1.00}
\draw[gp path] (2.199,0.405)--(2.199,0.495);
\node[gp node center] at (2.199,0.159) {2};
\gpcolor{color=gp lt color axes}
\gpsetlinetype{gp lt axes}
\gpsetdashtype{gp dt axes}
\gpsetlinewidth{0.50}
\draw[gp path] (2.960,0.405)--(2.960,2.431);
\gpcolor{color=gp lt color border}
\gpsetlinetype{gp lt border}
\gpsetdashtype{gp dt solid}
\gpsetlinewidth{1.00}
\draw[gp path] (2.960,0.405)--(2.960,0.495);
\node[gp node center] at (2.960,0.159) {3};
\gpcolor{color=gp lt color axes}
\gpsetlinetype{gp lt axes}
\gpsetdashtype{gp dt axes}
\gpsetlinewidth{0.50}
\draw[gp path] (3.721,0.405)--(3.721,2.431);
\gpcolor{color=gp lt color border}
\gpsetlinetype{gp lt border}
\gpsetdashtype{gp dt solid}
\gpsetlinewidth{1.00}
\draw[gp path] (3.721,0.405)--(3.721,0.495);
\node[gp node center] at (3.721,0.159) {4};
\draw[gp path] (3.721,0.405)--(3.631,0.405);
\draw[gp path] (3.721,0.912)--(3.631,0.912);
\draw[gp path] (3.721,1.418)--(3.631,1.418);
\draw[gp path] (3.721,1.925)--(3.631,1.925);
\draw[gp path] (3.721,2.431)--(3.631,2.431);
\draw[gp path] (0.676,2.431)--(0.676,0.405)--(3.721,0.405)--(3.721,2.431)--cycle;
\gpcolor{rgb color={0.471,0.369,0.941}}
\gpsetlinewidth{3.00}
\draw[gp path] (0.691,0.689)--(0.693,2.431);
\gpsetpointsize{4.00}
\gp3point{gp mark 5}{}{(0.691,0.689)}
\gpcolor{rgb color={0.863,0.149,0.498}}
\draw[gp path] (0.691,0.446)--(0.825,0.446)--(0.899,0.446)--(0.973,0.446)--(1.048,0.446)%
  --(1.122,0.446)--(1.716,0.456)--(2.311,0.466)--(2.906,0.496)--(3.497,0.668)--(3.471,1.347);
\gp3point{gp mark 7}{}{(0.691,0.446)}
\gp3point{gp mark 7}{}{(0.825,0.446)}
\gp3point{gp mark 7}{}{(0.899,0.446)}
\gp3point{gp mark 7}{}{(0.973,0.446)}
\gp3point{gp mark 7}{}{(1.048,0.446)}
\gp3point{gp mark 7}{}{(1.122,0.446)}
\gp3point{gp mark 7}{}{(1.716,0.456)}
\gp3point{gp mark 7}{}{(2.311,0.466)}
\gp3point{gp mark 7}{}{(2.906,0.496)}
\gp3point{gp mark 7}{}{(3.497,0.668)}
\gp3point{gp mark 7}{}{(3.471,1.347)}
\gpcolor{rgb color={0.392,0.561,1.000}}
\draw[gp path] (0.691,0.456)--(0.824,0.456)--(0.898,0.456)--(0.969,0.456)--(1.037,0.476)%
  --(1.040,2.431);
\gp3point{gp mark 11}{}{(0.691,0.456)}
\gp3point{gp mark 11}{}{(0.824,0.456)}
\gp3point{gp mark 11}{}{(0.898,0.456)}
\gp3point{gp mark 11}{}{(0.969,0.456)}
\gp3point{gp mark 11}{}{(1.037,0.476)}
\gpcolor{color=gp lt color border}
\gpsetlinewidth{1.00}
\draw[gp path] (0.676,2.431)--(0.676,0.405)--(3.721,0.405)--(3.721,2.431)--cycle;
\node[gp node center,rotate=-270.0] at (-0.218,1.418) {50p RTT [µs]};
\node[gp node center] at (2.198,-0.209) {Tput [MReq/s]};
\node[gp node center] at (2.198,2.677) {UDP 50p};
\gpdefrectangularnode{gp plot 2}{\pgfpoint{0.676cm}{0.405cm}}{\pgfpoint{3.721cm}{2.431cm}}
\gpcolor{color=gp lt color axes}
\gpsetlinetype{gp lt axes}
\gpsetdashtype{gp dt axes}
\gpsetlinewidth{0.50}
\draw[gp path] (4.905,0.405)--(7.949,0.405);
\gpcolor{color=gp lt color border}
\gpsetlinetype{gp lt border}
\gpsetdashtype{gp dt solid}
\gpsetlinewidth{1.00}
\draw[gp path] (4.905,0.405)--(4.995,0.405);
\node[gp node right] at (4.758,0.405) {$0$};
\gpcolor{color=gp lt color axes}
\gpsetlinetype{gp lt axes}
\gpsetdashtype{gp dt axes}
\gpsetlinewidth{0.50}
\draw[gp path] (4.905,0.912)--(7.949,0.912);
\gpcolor{color=gp lt color border}
\gpsetlinetype{gp lt border}
\gpsetdashtype{gp dt solid}
\gpsetlinewidth{1.00}
\draw[gp path] (4.905,0.912)--(4.995,0.912);
\node[gp node right] at (4.758,0.912) {$50$};
\gpcolor{color=gp lt color axes}
\gpsetlinetype{gp lt axes}
\gpsetdashtype{gp dt axes}
\gpsetlinewidth{0.50}
\draw[gp path] (4.905,1.418)--(7.949,1.418);
\gpcolor{color=gp lt color border}
\gpsetlinetype{gp lt border}
\gpsetdashtype{gp dt solid}
\gpsetlinewidth{1.00}
\draw[gp path] (4.905,1.418)--(4.995,1.418);
\node[gp node right] at (4.758,1.418) {$100$};
\gpcolor{color=gp lt color axes}
\gpsetlinetype{gp lt axes}
\gpsetdashtype{gp dt axes}
\gpsetlinewidth{0.50}
\draw[gp path] (4.905,1.925)--(7.949,1.925);
\gpcolor{color=gp lt color border}
\gpsetlinetype{gp lt border}
\gpsetdashtype{gp dt solid}
\gpsetlinewidth{1.00}
\draw[gp path] (4.905,1.925)--(4.995,1.925);
\node[gp node right] at (4.758,1.925) {$150$};
\gpcolor{color=gp lt color axes}
\gpsetlinetype{gp lt axes}
\gpsetdashtype{gp dt axes}
\gpsetlinewidth{0.50}
\draw[gp path] (4.905,2.431)--(7.949,2.431);
\gpcolor{color=gp lt color border}
\gpsetlinetype{gp lt border}
\gpsetdashtype{gp dt solid}
\gpsetlinewidth{1.00}
\draw[gp path] (4.905,2.431)--(4.995,2.431);
\node[gp node right] at (4.758,2.431) {$200$};
\gpcolor{color=gp lt color axes}
\gpsetlinetype{gp lt axes}
\gpsetdashtype{gp dt axes}
\gpsetlinewidth{0.50}
\draw[gp path] (4.905,0.405)--(4.905,2.431);
\gpcolor{color=gp lt color border}
\gpsetlinetype{gp lt border}
\gpsetdashtype{gp dt solid}
\gpsetlinewidth{1.00}
\draw[gp path] (4.905,0.405)--(4.905,0.495);
\node[gp node center] at (4.905,0.159) {0};
\gpcolor{color=gp lt color axes}
\gpsetlinetype{gp lt axes}
\gpsetdashtype{gp dt axes}
\gpsetlinewidth{0.50}
\draw[gp path] (5.666,0.405)--(5.666,2.431);
\gpcolor{color=gp lt color border}
\gpsetlinetype{gp lt border}
\gpsetdashtype{gp dt solid}
\gpsetlinewidth{1.00}
\draw[gp path] (5.666,0.405)--(5.666,0.495);
\node[gp node center] at (5.666,0.159) {1};
\gpcolor{color=gp lt color axes}
\gpsetlinetype{gp lt axes}
\gpsetdashtype{gp dt axes}
\gpsetlinewidth{0.50}
\draw[gp path] (6.427,0.405)--(6.427,2.431);
\gpcolor{color=gp lt color border}
\gpsetlinetype{gp lt border}
\gpsetdashtype{gp dt solid}
\gpsetlinewidth{1.00}
\draw[gp path] (6.427,0.405)--(6.427,0.495);
\node[gp node center] at (6.427,0.159) {2};
\gpcolor{color=gp lt color axes}
\gpsetlinetype{gp lt axes}
\gpsetdashtype{gp dt axes}
\gpsetlinewidth{0.50}
\draw[gp path] (7.188,0.405)--(7.188,2.431);
\gpcolor{color=gp lt color border}
\gpsetlinetype{gp lt border}
\gpsetdashtype{gp dt solid}
\gpsetlinewidth{1.00}
\draw[gp path] (7.188,0.405)--(7.188,0.495);
\node[gp node center] at (7.188,0.159) {3};
\gpcolor{color=gp lt color axes}
\gpsetlinetype{gp lt axes}
\gpsetdashtype{gp dt axes}
\gpsetlinewidth{0.50}
\draw[gp path] (7.949,0.405)--(7.949,2.431);
\gpcolor{color=gp lt color border}
\gpsetlinetype{gp lt border}
\gpsetdashtype{gp dt solid}
\gpsetlinewidth{1.00}
\draw[gp path] (7.949,0.405)--(7.949,0.495);
\node[gp node center] at (7.949,0.159) {4};
\draw[gp path] (7.949,0.405)--(7.859,0.405);
\draw[gp path] (7.949,0.912)--(7.859,0.912);
\draw[gp path] (7.949,1.418)--(7.859,1.418);
\draw[gp path] (7.949,1.925)--(7.859,1.925);
\draw[gp path] (7.949,2.431)--(7.859,2.431);
\draw[gp path] (4.905,2.431)--(4.905,0.405)--(7.949,0.405)--(7.949,2.431)--cycle;
\gpcolor{rgb color={0.471,0.369,0.941}}
\gpsetlinewidth{3.00}
\draw[gp path] (4.920,1.023)--(4.920,2.431);
\gp3point{gp mark 5}{}{(4.920,1.023)}
\gpcolor{rgb color={0.863,0.149,0.498}}
\draw[gp path] (4.920,0.496)--(5.054,0.506)--(5.128,0.516)--(5.202,0.506)--(5.276,0.506)%
  --(5.351,0.506)--(5.945,0.567)--(6.540,0.608)--(7.134,0.679)--(7.725,1.074)--(7.700,1.621);
\gp3point{gp mark 7}{}{(4.920,0.496)}
\gp3point{gp mark 7}{}{(5.054,0.506)}
\gp3point{gp mark 7}{}{(5.128,0.516)}
\gp3point{gp mark 7}{}{(5.202,0.506)}
\gp3point{gp mark 7}{}{(5.276,0.506)}
\gp3point{gp mark 7}{}{(5.351,0.506)}
\gp3point{gp mark 7}{}{(5.945,0.567)}
\gp3point{gp mark 7}{}{(6.540,0.608)}
\gp3point{gp mark 7}{}{(7.134,0.679)}
\gp3point{gp mark 7}{}{(7.725,1.074)}
\gp3point{gp mark 7}{}{(7.700,1.621)}
\gpcolor{rgb color={0.392,0.561,1.000}}
\draw[gp path] (4.920,0.466)--(5.053,0.527)--(5.126,0.547)--(5.198,0.567)--(5.266,0.820)%
  --(5.268,2.431);
\gp3point{gp mark 11}{}{(4.920,0.466)}
\gp3point{gp mark 11}{}{(5.053,0.527)}
\gp3point{gp mark 11}{}{(5.126,0.547)}
\gp3point{gp mark 11}{}{(5.198,0.567)}
\gp3point{gp mark 11}{}{(5.266,0.820)}
\gpcolor{color=gp lt color border}
\gpsetlinewidth{1.00}
\draw[gp path] (4.905,2.431)--(4.905,0.405)--(7.949,0.405)--(7.949,2.431)--cycle;
\node[gp node center,rotate=-270.0] at (4.010,1.418) {99p RTT [µs]};
\node[gp node center] at (6.427,-0.209) {Tput [MReq/s]};
\node[gp node center] at (6.427,2.677) {UDP 99p};
\gpdefrectangularnode{gp plot 3}{\pgfpoint{4.905cm}{0.405cm}}{\pgfpoint{7.949cm}{2.431cm}}
\end{tikzpicture}
\caption{UDP latency and throughput under load; \sys reaches
  7.9$\times$ Demikernel throughput.}%
\label{fig:udp_lat_tp}%
\end{figure}

\textbf{UDP.}
UDP provides the complementary message-based case and exposes per-packet
overheads with minimal protocol machinery. In this evaluation, we compare
against Linux and Demikernel UDP. We increase load and saturate the RPC echo
server and record 50th- and 99th-percentile RTT in
\autoref{fig:udp_lat_tp}.

\sys UDP saturates at 3.7 million requests per second, compared to
470k requests per second for Demikernel, an improvement of
7.9$\times$. Under low load, \sys also achieves lower latency, with a
50th-percentile RTT of 4\,\textmu s and a 99th-percentile RTT of
10\,\textmu s, compared to 5\,\textmu s and 12\,\textmu s for
Demikernel. Linux performs substantially worse than both systems and does
not approach comparable throughput or latency.

\subsection{Joint Compilation Shrinks the Programmability Tax}
\label{ssec:eval:joint_compilation}

We now isolate the residual cost of programmability and ask how much joint
compilation removes it. We compare three configurations: isolated JIT
compilation of the tenant handler (\texttt{bpf}), joint compilation with
infrastructure protocols (\texttt{comb}), and a hand-optimised native
implementation without eBPF (\texttt{nat}) as an upper bound. UDP shows the
small-protocol case, while TCP shows that the benefit grows substantially with
protocol complexity.

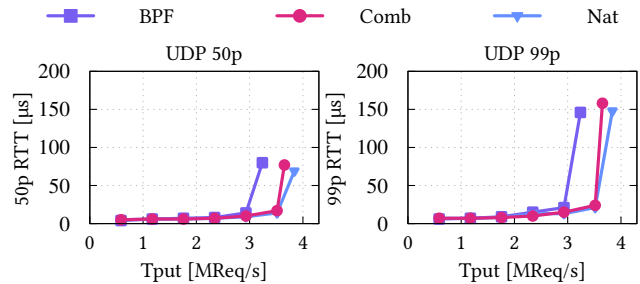
\begin{figure}%
\centering%
\begin{tikzpicture}[gnuplot]
\tikzset{every node/.append style={font={\fontsize{8.0pt}{9.6pt}\selectfont}}}
\path (0.000,0.000) rectangle (8.458,3.378);
\gpcolor{color=gp lt color border}
\gpcolor{rgb color={0.471,0.369,0.941}}
\gpsetlinewidth{1.00}
\gpsetpointsize{6.00}
\gp3point{gp mark 5}{}{(0.422,3.169)}
\gpcolor{color=gp lt color border}
\gpcolor{rgb color={0.863,0.149,0.498}}
\gp3point{gp mark 7}{}{(3.467,3.169)}
\gpcolor{color=gp lt color border}
\gpcolor{rgb color={0.392,0.561,1.000}}
\gp3point{gp mark 11}{}{(6.359,3.169)}
\gpcolor{color=gp lt color border}
\node[gp node center] at (1.564,3.169) {BPF};
\node[gp node center] at (4.609,3.169) {Comb};
\node[gp node center] at (7.501,3.169) {Nat};
\gpcolor{rgb color={0.471,0.369,0.941}}
\gpsetlinetype{gp lt border}
\gpsetdashtype{gp dt solid}
\gpsetlinewidth{3.00}
\draw[gp path](0.118,3.170)--(0.727,3.170);
\gpcolor{rgb color={0.863,0.149,0.498}}
\draw[gp path](3.163,3.170)--(3.772,3.170);
\gpcolor{rgb color={0.392,0.561,1.000}}
\draw[gp path](6.055,3.170)--(6.664,3.170);
\gpdefrectangularnode{gp plot 1}{\pgfpoint{0.422cm}{2.905cm}}{\pgfpoint{8.034cm}{3.377cm}}
\gpcolor{color=gp lt color axes}
\gpsetlinetype{gp lt axes}
\gpsetdashtype{gp dt axes}
\gpsetlinewidth{0.50}
\draw[gp path] (0.676,0.405)--(3.721,0.405);
\gpcolor{color=gp lt color border}
\gpsetlinetype{gp lt border}
\gpsetdashtype{gp dt solid}
\gpsetlinewidth{1.00}
\draw[gp path] (0.676,0.405)--(0.766,0.405);
\node[gp node right] at (0.529,0.405) {$0$};
\gpcolor{color=gp lt color axes}
\gpsetlinetype{gp lt axes}
\gpsetdashtype{gp dt axes}
\gpsetlinewidth{0.50}
\draw[gp path] (0.676,0.912)--(3.721,0.912);
\gpcolor{color=gp lt color border}
\gpsetlinetype{gp lt border}
\gpsetdashtype{gp dt solid}
\gpsetlinewidth{1.00}
\draw[gp path] (0.676,0.912)--(0.766,0.912);
\node[gp node right] at (0.529,0.912) {$50$};
\gpcolor{color=gp lt color axes}
\gpsetlinetype{gp lt axes}
\gpsetdashtype{gp dt axes}
\gpsetlinewidth{0.50}
\draw[gp path] (0.676,1.418)--(3.721,1.418);
\gpcolor{color=gp lt color border}
\gpsetlinetype{gp lt border}
\gpsetdashtype{gp dt solid}
\gpsetlinewidth{1.00}
\draw[gp path] (0.676,1.418)--(0.766,1.418);
\node[gp node right] at (0.529,1.418) {$100$};
\gpcolor{color=gp lt color axes}
\gpsetlinetype{gp lt axes}
\gpsetdashtype{gp dt axes}
\gpsetlinewidth{0.50}
\draw[gp path] (0.676,1.925)--(3.721,1.925);
\gpcolor{color=gp lt color border}
\gpsetlinetype{gp lt border}
\gpsetdashtype{gp dt solid}
\gpsetlinewidth{1.00}
\draw[gp path] (0.676,1.925)--(0.766,1.925);
\node[gp node right] at (0.529,1.925) {$150$};
\gpcolor{color=gp lt color axes}
\gpsetlinetype{gp lt axes}
\gpsetdashtype{gp dt axes}
\gpsetlinewidth{0.50}
\draw[gp path] (0.676,2.431)--(3.721,2.431);
\gpcolor{color=gp lt color border}
\gpsetlinetype{gp lt border}
\gpsetdashtype{gp dt solid}
\gpsetlinewidth{1.00}
\draw[gp path] (0.676,2.431)--(0.766,2.431);
\node[gp node right] at (0.529,2.431) {$200$};
\gpcolor{color=gp lt color axes}
\gpsetlinetype{gp lt axes}
\gpsetdashtype{gp dt axes}
\gpsetlinewidth{0.50}
\draw[gp path] (0.676,0.405)--(0.676,2.431);
\gpcolor{color=gp lt color border}
\gpsetlinetype{gp lt border}
\gpsetdashtype{gp dt solid}
\gpsetlinewidth{1.00}
\draw[gp path] (0.676,0.405)--(0.676,0.495);
\node[gp node center] at (0.676,0.159) {0};
\gpcolor{color=gp lt color axes}
\gpsetlinetype{gp lt axes}
\gpsetdashtype{gp dt axes}
\gpsetlinewidth{0.50}
\draw[gp path] (1.384,0.405)--(1.384,2.431);
\gpcolor{color=gp lt color border}
\gpsetlinetype{gp lt border}
\gpsetdashtype{gp dt solid}
\gpsetlinewidth{1.00}
\draw[gp path] (1.384,0.405)--(1.384,0.495);
\node[gp node center] at (1.384,0.159) {1};
\gpcolor{color=gp lt color axes}
\gpsetlinetype{gp lt axes}
\gpsetdashtype{gp dt axes}
\gpsetlinewidth{0.50}
\draw[gp path] (2.092,0.405)--(2.092,2.431);
\gpcolor{color=gp lt color border}
\gpsetlinetype{gp lt border}
\gpsetdashtype{gp dt solid}
\gpsetlinewidth{1.00}
\draw[gp path] (2.092,0.405)--(2.092,0.495);
\node[gp node center] at (2.092,0.159) {2};
\gpcolor{color=gp lt color axes}
\gpsetlinetype{gp lt axes}
\gpsetdashtype{gp dt axes}
\gpsetlinewidth{0.50}
\draw[gp path] (2.800,0.405)--(2.800,2.431);
\gpcolor{color=gp lt color border}
\gpsetlinetype{gp lt border}
\gpsetdashtype{gp dt solid}
\gpsetlinewidth{1.00}
\draw[gp path] (2.800,0.405)--(2.800,0.495);
\node[gp node center] at (2.800,0.159) {3};
\gpcolor{color=gp lt color axes}
\gpsetlinetype{gp lt axes}
\gpsetdashtype{gp dt axes}
\gpsetlinewidth{0.50}
\draw[gp path] (3.509,0.405)--(3.509,2.431);
\gpcolor{color=gp lt color border}
\gpsetlinetype{gp lt border}
\gpsetdashtype{gp dt solid}
\gpsetlinewidth{1.00}
\draw[gp path] (3.509,0.405)--(3.509,0.495);
\node[gp node center] at (3.509,0.159) {4};
\draw[gp path] (3.721,0.405)--(3.631,0.405);
\draw[gp path] (3.721,0.912)--(3.631,0.912);
\draw[gp path] (3.721,1.418)--(3.631,1.418);
\draw[gp path] (3.721,1.925)--(3.631,1.925);
\draw[gp path] (3.721,2.431)--(3.631,2.431);
\draw[gp path] (0.676,2.431)--(0.676,0.405)--(3.721,0.405)--(3.721,2.431)--cycle;
\gpcolor{rgb color={0.471,0.369,0.941}}
\gpsetlinewidth{3.00}
\draw[gp path] (1.091,0.446)--(1.506,0.466)--(1.921,0.476)--(2.335,0.486)--(2.751,0.547)%
  --(2.970,1.215);
\gpsetpointsize{4.00}
\gp3point{gp mark 5}{}{(1.091,0.446)}
\gp3point{gp mark 5}{}{(1.506,0.466)}
\gp3point{gp mark 5}{}{(1.921,0.476)}
\gp3point{gp mark 5}{}{(2.335,0.486)}
\gp3point{gp mark 5}{}{(2.751,0.547)}
\gp3point{gp mark 5}{}{(2.970,1.215)}
\gpcolor{rgb color={0.392,0.561,1.000}}
\draw[gp path] (1.091,0.456)--(1.505,0.466)--(1.921,0.466)--(2.336,0.476)--(2.751,0.496)%
  --(3.166,0.547)--(3.395,1.104);
\gp3point{gp mark 11}{}{(1.091,0.456)}
\gp3point{gp mark 11}{}{(1.505,0.466)}
\gp3point{gp mark 11}{}{(1.921,0.466)}
\gp3point{gp mark 11}{}{(2.336,0.476)}
\gp3point{gp mark 11}{}{(2.751,0.496)}
\gp3point{gp mark 11}{}{(3.166,0.547)}
\gp3point{gp mark 11}{}{(3.395,1.104)}
\gpcolor{rgb color={0.863,0.149,0.498}}
\draw[gp path] (1.091,0.456)--(1.506,0.466)--(1.921,0.466)--(2.336,0.476)--(2.751,0.506)%
  --(3.165,0.577)--(3.263,1.185);
\gp3point{gp mark 7}{}{(1.091,0.456)}
\gp3point{gp mark 7}{}{(1.506,0.466)}
\gp3point{gp mark 7}{}{(1.921,0.466)}
\gp3point{gp mark 7}{}{(2.336,0.476)}
\gp3point{gp mark 7}{}{(2.751,0.506)}
\gp3point{gp mark 7}{}{(3.165,0.577)}
\gp3point{gp mark 7}{}{(3.263,1.185)}
\gpcolor{color=gp lt color border}
\gpsetlinewidth{1.00}
\draw[gp path] (0.676,2.431)--(0.676,0.405)--(3.721,0.405)--(3.721,2.431)--cycle;
\node[gp node center,rotate=-270.0] at (-0.218,1.418) {50p RTT [µs]};
\node[gp node center] at (2.198,-0.209) {Tput [MReq/s]};
\node[gp node center] at (2.198,2.677) {UDP 50p};
\gpdefrectangularnode{gp plot 2}{\pgfpoint{0.676cm}{0.405cm}}{\pgfpoint{3.721cm}{2.431cm}}
\gpcolor{color=gp lt color axes}
\gpsetlinetype{gp lt axes}
\gpsetdashtype{gp dt axes}
\gpsetlinewidth{0.50}
\draw[gp path] (4.905,0.405)--(7.949,0.405);
\gpcolor{color=gp lt color border}
\gpsetlinetype{gp lt border}
\gpsetdashtype{gp dt solid}
\gpsetlinewidth{1.00}
\draw[gp path] (4.905,0.405)--(4.995,0.405);
\node[gp node right] at (4.758,0.405) {$0$};
\gpcolor{color=gp lt color axes}
\gpsetlinetype{gp lt axes}
\gpsetdashtype{gp dt axes}
\gpsetlinewidth{0.50}
\draw[gp path] (4.905,0.912)--(7.949,0.912);
\gpcolor{color=gp lt color border}
\gpsetlinetype{gp lt border}
\gpsetdashtype{gp dt solid}
\gpsetlinewidth{1.00}
\draw[gp path] (4.905,0.912)--(4.995,0.912);
\node[gp node right] at (4.758,0.912) {$50$};
\gpcolor{color=gp lt color axes}
\gpsetlinetype{gp lt axes}
\gpsetdashtype{gp dt axes}
\gpsetlinewidth{0.50}
\draw[gp path] (4.905,1.418)--(7.949,1.418);
\gpcolor{color=gp lt color border}
\gpsetlinetype{gp lt border}
\gpsetdashtype{gp dt solid}
\gpsetlinewidth{1.00}
\draw[gp path] (4.905,1.418)--(4.995,1.418);
\node[gp node right] at (4.758,1.418) {$100$};
\gpcolor{color=gp lt color axes}
\gpsetlinetype{gp lt axes}
\gpsetdashtype{gp dt axes}
\gpsetlinewidth{0.50}
\draw[gp path] (4.905,1.925)--(7.949,1.925);
\gpcolor{color=gp lt color border}
\gpsetlinetype{gp lt border}
\gpsetdashtype{gp dt solid}
\gpsetlinewidth{1.00}
\draw[gp path] (4.905,1.925)--(4.995,1.925);
\node[gp node right] at (4.758,1.925) {$150$};
\gpcolor{color=gp lt color axes}
\gpsetlinetype{gp lt axes}
\gpsetdashtype{gp dt axes}
\gpsetlinewidth{0.50}
\draw[gp path] (4.905,2.431)--(7.949,2.431);
\gpcolor{color=gp lt color border}
\gpsetlinetype{gp lt border}
\gpsetdashtype{gp dt solid}
\gpsetlinewidth{1.00}
\draw[gp path] (4.905,2.431)--(4.995,2.431);
\node[gp node right] at (4.758,2.431) {$200$};
\gpcolor{color=gp lt color axes}
\gpsetlinetype{gp lt axes}
\gpsetdashtype{gp dt axes}
\gpsetlinewidth{0.50}
\draw[gp path] (4.905,0.405)--(4.905,2.431);
\gpcolor{color=gp lt color border}
\gpsetlinetype{gp lt border}
\gpsetdashtype{gp dt solid}
\gpsetlinewidth{1.00}
\draw[gp path] (4.905,0.405)--(4.905,0.495);
\node[gp node center] at (4.905,0.159) {0};
\gpcolor{color=gp lt color axes}
\gpsetlinetype{gp lt axes}
\gpsetdashtype{gp dt axes}
\gpsetlinewidth{0.50}
\draw[gp path] (5.613,0.405)--(5.613,2.431);
\gpcolor{color=gp lt color border}
\gpsetlinetype{gp lt border}
\gpsetdashtype{gp dt solid}
\gpsetlinewidth{1.00}
\draw[gp path] (5.613,0.405)--(5.613,0.495);
\node[gp node center] at (5.613,0.159) {1};
\gpcolor{color=gp lt color axes}
\gpsetlinetype{gp lt axes}
\gpsetdashtype{gp dt axes}
\gpsetlinewidth{0.50}
\draw[gp path] (6.321,0.405)--(6.321,2.431);
\gpcolor{color=gp lt color border}
\gpsetlinetype{gp lt border}
\gpsetdashtype{gp dt solid}
\gpsetlinewidth{1.00}
\draw[gp path] (6.321,0.405)--(6.321,0.495);
\node[gp node center] at (6.321,0.159) {2};
\gpcolor{color=gp lt color axes}
\gpsetlinetype{gp lt axes}
\gpsetdashtype{gp dt axes}
\gpsetlinewidth{0.50}
\draw[gp path] (7.029,0.405)--(7.029,2.431);
\gpcolor{color=gp lt color border}
\gpsetlinetype{gp lt border}
\gpsetdashtype{gp dt solid}
\gpsetlinewidth{1.00}
\draw[gp path] (7.029,0.405)--(7.029,0.495);
\node[gp node center] at (7.029,0.159) {3};
\gpcolor{color=gp lt color axes}
\gpsetlinetype{gp lt axes}
\gpsetdashtype{gp dt axes}
\gpsetlinewidth{0.50}
\draw[gp path] (7.737,0.405)--(7.737,2.431);
\gpcolor{color=gp lt color border}
\gpsetlinetype{gp lt border}
\gpsetdashtype{gp dt solid}
\gpsetlinewidth{1.00}
\draw[gp path] (7.737,0.405)--(7.737,0.495);
\node[gp node center] at (7.737,0.159) {4};
\draw[gp path] (7.949,0.405)--(7.859,0.405);
\draw[gp path] (7.949,0.912)--(7.859,0.912);
\draw[gp path] (7.949,1.418)--(7.859,1.418);
\draw[gp path] (7.949,1.925)--(7.859,1.925);
\draw[gp path] (7.949,2.431)--(7.859,2.431);
\draw[gp path] (4.905,2.431)--(4.905,0.405)--(7.949,0.405)--(7.949,2.431)--cycle;
\gpcolor{rgb color={0.471,0.369,0.941}}
\gpsetlinewidth{3.00}
\draw[gp path] (5.320,0.466)--(5.735,0.476)--(6.149,0.496)--(6.564,0.557)--(6.979,0.618)%
  --(7.198,1.884);
\gp3point{gp mark 5}{}{(5.320,0.466)}
\gp3point{gp mark 5}{}{(5.735,0.476)}
\gp3point{gp mark 5}{}{(6.149,0.496)}
\gp3point{gp mark 5}{}{(6.564,0.557)}
\gp3point{gp mark 5}{}{(6.979,0.618)}
\gp3point{gp mark 5}{}{(7.198,1.884)}
\gpcolor{rgb color={0.392,0.561,1.000}}
\draw[gp path] (5.320,0.476)--(5.734,0.476)--(6.149,0.486)--(6.564,0.537)--(6.979,0.537)%
  --(7.394,0.618)--(7.623,1.904);
\gp3point{gp mark 11}{}{(5.320,0.476)}
\gp3point{gp mark 11}{}{(5.734,0.476)}
\gp3point{gp mark 11}{}{(6.149,0.486)}
\gp3point{gp mark 11}{}{(6.564,0.537)}
\gp3point{gp mark 11}{}{(6.979,0.537)}
\gp3point{gp mark 11}{}{(7.394,0.618)}
\gp3point{gp mark 11}{}{(7.623,1.904)}
\gpcolor{rgb color={0.863,0.149,0.498}}
\draw[gp path] (5.320,0.476)--(5.735,0.476)--(6.149,0.486)--(6.564,0.506)--(6.979,0.557)%
  --(7.394,0.648)--(7.491,2.006);
\gp3point{gp mark 7}{}{(5.320,0.476)}
\gp3point{gp mark 7}{}{(5.735,0.476)}
\gp3point{gp mark 7}{}{(6.149,0.486)}
\gp3point{gp mark 7}{}{(6.564,0.506)}
\gp3point{gp mark 7}{}{(6.979,0.557)}
\gp3point{gp mark 7}{}{(7.394,0.648)}
\gp3point{gp mark 7}{}{(7.491,2.006)}
\gpcolor{color=gp lt color border}
\gpsetlinewidth{1.00}
\draw[gp path] (4.905,2.431)--(4.905,0.405)--(7.949,0.405)--(7.949,2.431)--cycle;
\node[gp node center,rotate=-270.0] at (3.936,1.418) {99p RTT [µs]};
\node[gp node center] at (6.427,-0.209) {Tput [MReq/s]};
\node[gp node center] at (6.427,2.677) {UDP 99p};
\gpdefrectangularnode{gp plot 3}{\pgfpoint{4.905cm}{0.405cm}}{\pgfpoint{7.949cm}{2.431cm}}
\end{tikzpicture}
\caption{Joint compilation modestly reduces the UDP programmability tax.}%
\label{fig:udp_jit}%
\end{figure}

\textbf{UDP.}
Under UDP, \autoref{fig:udp_jit}, the three configurations track each other
closely until the knee of the curve, after which latency rises sharply as the
system saturates. Joint compilation (\texttt{comb}) reaches this point at a
throughput 4.9\% below the native upper bound (\texttt{nat}), while isolated
JIT compilation of the tenant handler (\texttt{bpf}) saturates earlier with a
14.7\% slowdown to \texttt{nat}. The 50th- and 99th-percentile RTT follow the
same ordering, so for UDP most of the programmability tax appears as an
earlier saturation point rather than as a large low-load RTT gap.

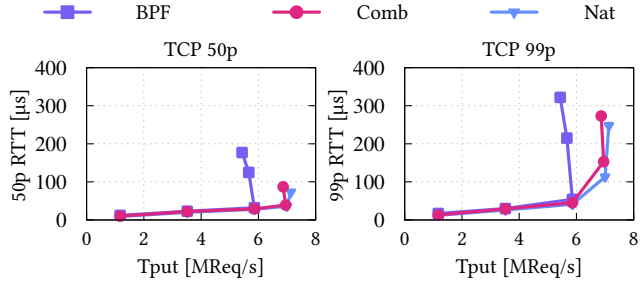
\begin{figure}%
\centering%
\begin{tikzpicture}[gnuplot]
\tikzset{every node/.append style={font={\fontsize{8.0pt}{9.6pt}\selectfont}}}
\path (0.000,0.000) rectangle (8.458,3.378);
\gpcolor{color=gp lt color border}
\gpcolor{rgb color={0.471,0.369,0.941}}
\gpsetlinewidth{1.00}
\gpsetpointsize{6.00}
\gp3point{gp mark 5}{}{(0.422,3.169)}
\gpcolor{color=gp lt color border}
\gpcolor{rgb color={0.863,0.149,0.498}}
\gp3point{gp mark 7}{}{(3.467,3.169)}
\gpcolor{color=gp lt color border}
\gpcolor{rgb color={0.392,0.561,1.000}}
\gp3point{gp mark 11}{}{(6.359,3.169)}
\gpcolor{color=gp lt color border}
\node[gp node center] at (1.564,3.169) {BPF};
\node[gp node center] at (4.609,3.169) {Comb};
\node[gp node center] at (7.501,3.169) {Nat};
\gpcolor{rgb color={0.471,0.369,0.941}}
\gpsetlinetype{gp lt border}
\gpsetdashtype{gp dt solid}
\gpsetlinewidth{3.00}
\draw[gp path](0.118,3.170)--(0.727,3.170);
\gpcolor{rgb color={0.863,0.149,0.498}}
\draw[gp path](3.163,3.170)--(3.772,3.170);
\gpcolor{rgb color={0.392,0.561,1.000}}
\draw[gp path](6.055,3.170)--(6.664,3.170);
\gpdefrectangularnode{gp plot 1}{\pgfpoint{0.422cm}{2.905cm}}{\pgfpoint{8.034cm}{3.377cm}}
\gpcolor{color=gp lt color axes}
\gpsetlinetype{gp lt axes}
\gpsetdashtype{gp dt axes}
\gpsetlinewidth{0.50}
\draw[gp path] (0.676,0.405)--(3.721,0.405);
\gpcolor{color=gp lt color border}
\gpsetlinetype{gp lt border}
\gpsetdashtype{gp dt solid}
\gpsetlinewidth{1.00}
\draw[gp path] (0.676,0.405)--(0.766,0.405);
\node[gp node right] at (0.529,0.405) {$0$};
\gpcolor{color=gp lt color axes}
\gpsetlinetype{gp lt axes}
\gpsetdashtype{gp dt axes}
\gpsetlinewidth{0.50}
\draw[gp path] (0.676,0.912)--(3.721,0.912);
\gpcolor{color=gp lt color border}
\gpsetlinetype{gp lt border}
\gpsetdashtype{gp dt solid}
\gpsetlinewidth{1.00}
\draw[gp path] (0.676,0.912)--(0.766,0.912);
\node[gp node right] at (0.529,0.912) {$100$};
\gpcolor{color=gp lt color axes}
\gpsetlinetype{gp lt axes}
\gpsetdashtype{gp dt axes}
\gpsetlinewidth{0.50}
\draw[gp path] (0.676,1.418)--(3.721,1.418);
\gpcolor{color=gp lt color border}
\gpsetlinetype{gp lt border}
\gpsetdashtype{gp dt solid}
\gpsetlinewidth{1.00}
\draw[gp path] (0.676,1.418)--(0.766,1.418);
\node[gp node right] at (0.529,1.418) {$200$};
\gpcolor{color=gp lt color axes}
\gpsetlinetype{gp lt axes}
\gpsetdashtype{gp dt axes}
\gpsetlinewidth{0.50}
\draw[gp path] (0.676,1.925)--(3.721,1.925);
\gpcolor{color=gp lt color border}
\gpsetlinetype{gp lt border}
\gpsetdashtype{gp dt solid}
\gpsetlinewidth{1.00}
\draw[gp path] (0.676,1.925)--(0.766,1.925);
\node[gp node right] at (0.529,1.925) {$300$};
\gpcolor{color=gp lt color axes}
\gpsetlinetype{gp lt axes}
\gpsetdashtype{gp dt axes}
\gpsetlinewidth{0.50}
\draw[gp path] (0.676,2.431)--(3.721,2.431);
\gpcolor{color=gp lt color border}
\gpsetlinetype{gp lt border}
\gpsetdashtype{gp dt solid}
\gpsetlinewidth{1.00}
\draw[gp path] (0.676,2.431)--(0.766,2.431);
\node[gp node right] at (0.529,2.431) {$400$};
\gpcolor{color=gp lt color axes}
\gpsetlinetype{gp lt axes}
\gpsetdashtype{gp dt axes}
\gpsetlinewidth{0.50}
\draw[gp path] (0.676,0.405)--(0.676,2.431);
\gpcolor{color=gp lt color border}
\gpsetlinetype{gp lt border}
\gpsetdashtype{gp dt solid}
\gpsetlinewidth{1.00}
\draw[gp path] (0.676,0.405)--(0.676,0.495);
\node[gp node center] at (0.676,0.159) {0};
\gpcolor{color=gp lt color axes}
\gpsetlinetype{gp lt axes}
\gpsetdashtype{gp dt axes}
\gpsetlinewidth{0.50}
\draw[gp path] (1.437,0.405)--(1.437,2.431);
\gpcolor{color=gp lt color border}
\gpsetlinetype{gp lt border}
\gpsetdashtype{gp dt solid}
\gpsetlinewidth{1.00}
\draw[gp path] (1.437,0.405)--(1.437,0.495);
\node[gp node center] at (1.437,0.159) {2};
\gpcolor{color=gp lt color axes}
\gpsetlinetype{gp lt axes}
\gpsetdashtype{gp dt axes}
\gpsetlinewidth{0.50}
\draw[gp path] (2.199,0.405)--(2.199,2.431);
\gpcolor{color=gp lt color border}
\gpsetlinetype{gp lt border}
\gpsetdashtype{gp dt solid}
\gpsetlinewidth{1.00}
\draw[gp path] (2.199,0.405)--(2.199,0.495);
\node[gp node center] at (2.199,0.159) {4};
\gpcolor{color=gp lt color axes}
\gpsetlinetype{gp lt axes}
\gpsetdashtype{gp dt axes}
\gpsetlinewidth{0.50}
\draw[gp path] (2.960,0.405)--(2.960,2.431);
\gpcolor{color=gp lt color border}
\gpsetlinetype{gp lt border}
\gpsetdashtype{gp dt solid}
\gpsetlinewidth{1.00}
\draw[gp path] (2.960,0.405)--(2.960,0.495);
\node[gp node center] at (2.960,0.159) {6};
\gpcolor{color=gp lt color axes}
\gpsetlinetype{gp lt axes}
\gpsetdashtype{gp dt axes}
\gpsetlinewidth{0.50}
\draw[gp path] (3.721,0.405)--(3.721,2.431);
\gpcolor{color=gp lt color border}
\gpsetlinetype{gp lt border}
\gpsetdashtype{gp dt solid}
\gpsetlinewidth{1.00}
\draw[gp path] (3.721,0.405)--(3.721,0.495);
\node[gp node center] at (3.721,0.159) {8};
\draw[gp path] (3.721,0.405)--(3.631,0.405);
\draw[gp path] (3.721,0.912)--(3.631,0.912);
\draw[gp path] (3.721,1.418)--(3.631,1.418);
\draw[gp path] (3.721,1.925)--(3.631,1.925);
\draw[gp path] (3.721,2.431)--(3.631,2.431);
\draw[gp path] (0.676,2.431)--(0.676,0.405)--(3.721,0.405)--(3.721,2.431)--cycle;
\gpcolor{rgb color={0.471,0.369,0.941}}
\gpsetlinewidth{3.00}
\draw[gp path] (1.122,0.466)--(2.014,0.521)--(2.906,0.567)--(2.832,1.038)--(2.743,1.302);
\gpsetpointsize{4.00}
\gp3point{gp mark 5}{}{(1.122,0.466)}
\gp3point{gp mark 5}{}{(2.014,0.521)}
\gp3point{gp mark 5}{}{(2.906,0.567)}
\gp3point{gp mark 5}{}{(2.832,1.038)}
\gp3point{gp mark 5}{}{(2.743,1.302)}
\gpcolor{rgb color={0.392,0.561,1.000}}
\draw[gp path] (1.122,0.451)--(2.014,0.506)--(2.906,0.547)--(3.340,0.587)--(3.391,0.775);
\gp3point{gp mark 11}{}{(1.122,0.451)}
\gp3point{gp mark 11}{}{(2.014,0.506)}
\gp3point{gp mark 11}{}{(2.906,0.547)}
\gp3point{gp mark 11}{}{(3.340,0.587)}
\gp3point{gp mark 11}{}{(3.391,0.775)}
\gpcolor{rgb color={0.863,0.149,0.498}}
\draw[gp path] (1.122,0.456)--(2.014,0.516)--(2.906,0.552)--(3.322,0.603)--(3.289,0.846);
\gp3point{gp mark 7}{}{(1.122,0.456)}
\gp3point{gp mark 7}{}{(2.014,0.516)}
\gp3point{gp mark 7}{}{(2.906,0.552)}
\gp3point{gp mark 7}{}{(3.322,0.603)}
\gp3point{gp mark 7}{}{(3.289,0.846)}
\gpcolor{color=gp lt color border}
\gpsetlinewidth{1.00}
\draw[gp path] (0.676,2.431)--(0.676,0.405)--(3.721,0.405)--(3.721,2.431)--cycle;
\node[gp node center,rotate=-270.0] at (-0.218,1.418) {50p RTT [µs]};
\node[gp node center] at (2.198,-0.209) {Tput [MReq/s]};
\node[gp node center] at (2.198,2.677) {TCP 50p};
\gpdefrectangularnode{gp plot 2}{\pgfpoint{0.676cm}{0.405cm}}{\pgfpoint{3.721cm}{2.431cm}}
\gpcolor{color=gp lt color axes}
\gpsetlinetype{gp lt axes}
\gpsetdashtype{gp dt axes}
\gpsetlinewidth{0.50}
\draw[gp path] (4.905,0.405)--(7.949,0.405);
\gpcolor{color=gp lt color border}
\gpsetlinetype{gp lt border}
\gpsetdashtype{gp dt solid}
\gpsetlinewidth{1.00}
\draw[gp path] (4.905,0.405)--(4.995,0.405);
\node[gp node right] at (4.758,0.405) {$0$};
\gpcolor{color=gp lt color axes}
\gpsetlinetype{gp lt axes}
\gpsetdashtype{gp dt axes}
\gpsetlinewidth{0.50}
\draw[gp path] (4.905,0.912)--(7.949,0.912);
\gpcolor{color=gp lt color border}
\gpsetlinetype{gp lt border}
\gpsetdashtype{gp dt solid}
\gpsetlinewidth{1.00}
\draw[gp path] (4.905,0.912)--(4.995,0.912);
\node[gp node right] at (4.758,0.912) {$100$};
\gpcolor{color=gp lt color axes}
\gpsetlinetype{gp lt axes}
\gpsetdashtype{gp dt axes}
\gpsetlinewidth{0.50}
\draw[gp path] (4.905,1.418)--(7.949,1.418);
\gpcolor{color=gp lt color border}
\gpsetlinetype{gp lt border}
\gpsetdashtype{gp dt solid}
\gpsetlinewidth{1.00}
\draw[gp path] (4.905,1.418)--(4.995,1.418);
\node[gp node right] at (4.758,1.418) {$200$};
\gpcolor{color=gp lt color axes}
\gpsetlinetype{gp lt axes}
\gpsetdashtype{gp dt axes}
\gpsetlinewidth{0.50}
\draw[gp path] (4.905,1.925)--(7.949,1.925);
\gpcolor{color=gp lt color border}
\gpsetlinetype{gp lt border}
\gpsetdashtype{gp dt solid}
\gpsetlinewidth{1.00}
\draw[gp path] (4.905,1.925)--(4.995,1.925);
\node[gp node right] at (4.758,1.925) {$300$};
\gpcolor{color=gp lt color axes}
\gpsetlinetype{gp lt axes}
\gpsetdashtype{gp dt axes}
\gpsetlinewidth{0.50}
\draw[gp path] (4.905,2.431)--(7.949,2.431);
\gpcolor{color=gp lt color border}
\gpsetlinetype{gp lt border}
\gpsetdashtype{gp dt solid}
\gpsetlinewidth{1.00}
\draw[gp path] (4.905,2.431)--(4.995,2.431);
\node[gp node right] at (4.758,2.431) {$400$};
\gpcolor{color=gp lt color axes}
\gpsetlinetype{gp lt axes}
\gpsetdashtype{gp dt axes}
\gpsetlinewidth{0.50}
\draw[gp path] (4.905,0.405)--(4.905,2.431);
\gpcolor{color=gp lt color border}
\gpsetlinetype{gp lt border}
\gpsetdashtype{gp dt solid}
\gpsetlinewidth{1.00}
\draw[gp path] (4.905,0.405)--(4.905,0.495);
\node[gp node center] at (4.905,0.159) {0};
\gpcolor{color=gp lt color axes}
\gpsetlinetype{gp lt axes}
\gpsetdashtype{gp dt axes}
\gpsetlinewidth{0.50}
\draw[gp path] (5.666,0.405)--(5.666,2.431);
\gpcolor{color=gp lt color border}
\gpsetlinetype{gp lt border}
\gpsetdashtype{gp dt solid}
\gpsetlinewidth{1.00}
\draw[gp path] (5.666,0.405)--(5.666,0.495);
\node[gp node center] at (5.666,0.159) {2};
\gpcolor{color=gp lt color axes}
\gpsetlinetype{gp lt axes}
\gpsetdashtype{gp dt axes}
\gpsetlinewidth{0.50}
\draw[gp path] (6.427,0.405)--(6.427,2.431);
\gpcolor{color=gp lt color border}
\gpsetlinetype{gp lt border}
\gpsetdashtype{gp dt solid}
\gpsetlinewidth{1.00}
\draw[gp path] (6.427,0.405)--(6.427,0.495);
\node[gp node center] at (6.427,0.159) {4};
\gpcolor{color=gp lt color axes}
\gpsetlinetype{gp lt axes}
\gpsetdashtype{gp dt axes}
\gpsetlinewidth{0.50}
\draw[gp path] (7.188,0.405)--(7.188,2.431);
\gpcolor{color=gp lt color border}
\gpsetlinetype{gp lt border}
\gpsetdashtype{gp dt solid}
\gpsetlinewidth{1.00}
\draw[gp path] (7.188,0.405)--(7.188,0.495);
\node[gp node center] at (7.188,0.159) {6};
\gpcolor{color=gp lt color axes}
\gpsetlinetype{gp lt axes}
\gpsetdashtype{gp dt axes}
\gpsetlinewidth{0.50}
\draw[gp path] (7.949,0.405)--(7.949,2.431);
\gpcolor{color=gp lt color border}
\gpsetlinetype{gp lt border}
\gpsetdashtype{gp dt solid}
\gpsetlinewidth{1.00}
\draw[gp path] (7.949,0.405)--(7.949,0.495);
\node[gp node center] at (7.949,0.159) {8};
\draw[gp path] (7.949,0.405)--(7.859,0.405);
\draw[gp path] (7.949,0.912)--(7.859,0.912);
\draw[gp path] (7.949,1.418)--(7.859,1.418);
\draw[gp path] (7.949,1.925)--(7.859,1.925);
\draw[gp path] (7.949,2.431)--(7.859,2.431);
\draw[gp path] (4.905,2.431)--(4.905,0.405)--(7.949,0.405)--(7.949,2.431)--cycle;
\gpcolor{rgb color={0.471,0.369,0.941}}
\gpsetlinewidth{3.00}
\draw[gp path] (5.351,0.491)--(6.243,0.557)--(7.134,0.679)--(7.061,1.494)--(6.972,2.036);
\gp3point{gp mark 5}{}{(5.351,0.491)}
\gp3point{gp mark 5}{}{(6.243,0.557)}
\gp3point{gp mark 5}{}{(7.134,0.679)}
\gp3point{gp mark 5}{}{(7.061,1.494)}
\gp3point{gp mark 5}{}{(6.972,2.036)}
\gpcolor{rgb color={0.392,0.561,1.000}}
\draw[gp path] (5.351,0.466)--(6.243,0.537)--(7.134,0.613)--(7.568,0.977)--(7.619,1.661);
\gp3point{gp mark 11}{}{(5.351,0.466)}
\gp3point{gp mark 11}{}{(6.243,0.537)}
\gp3point{gp mark 11}{}{(7.134,0.613)}
\gp3point{gp mark 11}{}{(7.568,0.977)}
\gp3point{gp mark 11}{}{(7.619,1.661)}
\gpcolor{rgb color={0.863,0.149,0.498}}
\draw[gp path] (5.351,0.471)--(6.243,0.552)--(7.134,0.633)--(7.550,1.180)--(7.517,1.788);
\gp3point{gp mark 7}{}{(5.351,0.471)}
\gp3point{gp mark 7}{}{(6.243,0.552)}
\gp3point{gp mark 7}{}{(7.134,0.633)}
\gp3point{gp mark 7}{}{(7.550,1.180)}
\gp3point{gp mark 7}{}{(7.517,1.788)}
\gpcolor{color=gp lt color border}
\gpsetlinewidth{1.00}
\draw[gp path] (4.905,2.431)--(4.905,0.405)--(7.949,0.405)--(7.949,2.431)--cycle;
\node[gp node center,rotate=-270.0] at (4.010,1.418) {99p RTT [µs]};
\node[gp node center] at (6.427,-0.209) {Tput [MReq/s]};
\node[gp node center] at (6.427,2.677) {TCP 99p};
\gpdefrectangularnode{gp plot 3}{\pgfpoint{4.905cm}{0.405cm}}{\pgfpoint{7.949cm}{2.431cm}}
\end{tikzpicture}
\caption{Joint compilation cuts the TCP programmability tax from 23.9\% to
  3.8\%.}%
\label{fig:tcp_jit}%
\end{figure}

\textbf{TCP.}
In \autoref{fig:tcp_jit}, joint compilation (\texttt{comb}) achieves
near-native performance. Its 99th-percentile RTT reaches the saturation knee
at throughput only 3.8\% lower than the hand-optimised native implementation
(\texttt{nat}), whereas
isolated JIT compilation of the tenant handler (\texttt{bpf}) incurs a 23.9\%
slowdown. The 50th-percentile RTT follows the same trend and exhibits the same
relative slowdowns. Compared to UDP, TCP shows a much larger programmability
tax without joint compilation and a much larger recovery with it. Joint
compilation therefore matters most for complex protocols, where abstraction
boundaries would otherwise leave substantial optimisation headroom unused.

\subsection{Tenant-Defined Protocols Preserve Shared-Stack Isolation}
\label{ssec:eval:isolation}

Prior work has shown fine-grained accounting and scheduling can isolate a
fixed shared host datapath~\cite{stolet:virtuoso}. The question for \sys is
whether that isolation survives tenant-defined fast path handlers, whose cost
varies across protocols and packets. We therefore deploy a latency-sensitive
UDP tenant as the victim and a full TCP tenant as the adversary, then increase
the adversary's load by scaling its number of application threads to create
contention in the shared datapath. We measure the victim's 99th-percentile RTT and
the throughput of both tenants. We choose TCP as the adversary because its
stateful and complex processing is a realistic worst case for a
tenant-programmable shared fast path.

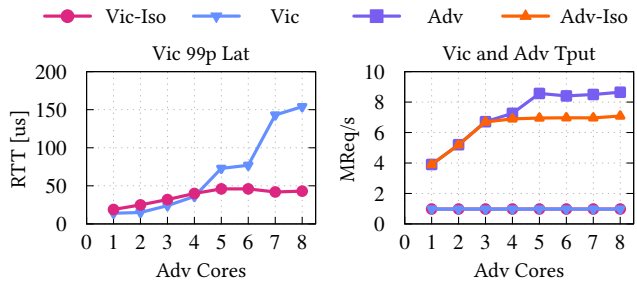
\begin{figure}%
\centering%
\begin{tikzpicture}[gnuplot]
\tikzset{every node/.append style={font={\fontsize{8.0pt}{9.6pt}\selectfont}}}
\path (0.000,0.000) rectangle (8.458,3.378);
\gpcolor{color=gp lt color border}
\gpcolor{rgb color={0.863,0.149,0.498}}
\gpsetlinewidth{1.00}
\gpsetpointsize{6.00}
\gp3point{gp mark 7}{}{(0.422,3.169)}
\gpcolor{color=gp lt color border}
\gpcolor{rgb color={0.392,0.561,1.000}}
\gp3point{gp mark 11}{}{(2.401,3.169)}
\gpcolor{color=gp lt color border}
\gpcolor{rgb color={0.471,0.369,0.941}}
\gp3point{gp mark 5}{}{(4.532,3.169)}
\gpcolor{color=gp lt color border}
\gpcolor{rgb color={0.996,0.380,0.000}}
\gp3point{gp mark 9}{}{(6.512,3.169)}
\gpcolor{color=gp lt color border}
\node[gp node center] at (1.335,3.169) {Vic-Iso};
\node[gp node center] at (3.315,3.169) {Vic};
\node[gp node center] at (5.446,3.169) {Adv};
\node[gp node center] at (7.425,3.169) {Adv-Iso};
\gpcolor{rgb color={0.863,0.149,0.498}}
\gpsetlinetype{gp lt border}
\gpsetdashtype{gp dt solid}
\gpsetlinewidth{3.00}
\draw[gp path](0.118,3.170)--(0.727,3.170);
\gpcolor{rgb color={0.392,0.561,1.000}}
\draw[gp path](2.097,3.170)--(2.706,3.170);
\gpcolor{rgb color={0.471,0.369,0.941}}
\draw[gp path](4.229,3.170)--(4.837,3.170);
\gpcolor{rgb color={0.996,0.380,0.000}}
\draw[gp path](6.208,3.170)--(6.817,3.170);
\gpdefrectangularnode{gp plot 1}{\pgfpoint{0.422cm}{2.905cm}}{\pgfpoint{8.034cm}{3.377cm}}
\gpcolor{color=gp lt color axes}
\gpsetlinetype{gp lt axes}
\gpsetdashtype{gp dt axes}
\gpsetlinewidth{0.50}
\draw[gp path] (0.676,0.405)--(3.721,0.405);
\gpcolor{color=gp lt color border}
\gpsetlinetype{gp lt border}
\gpsetdashtype{gp dt solid}
\gpsetlinewidth{1.00}
\draw[gp path] (0.676,0.405)--(0.766,0.405);
\node[gp node right] at (0.529,0.405) {$0$};
\gpcolor{color=gp lt color axes}
\gpsetlinetype{gp lt axes}
\gpsetdashtype{gp dt axes}
\gpsetlinewidth{0.50}
\draw[gp path] (0.676,0.912)--(3.721,0.912);
\gpcolor{color=gp lt color border}
\gpsetlinetype{gp lt border}
\gpsetdashtype{gp dt solid}
\gpsetlinewidth{1.00}
\draw[gp path] (0.676,0.912)--(0.766,0.912);
\node[gp node right] at (0.529,0.912) {$50$};
\gpcolor{color=gp lt color axes}
\gpsetlinetype{gp lt axes}
\gpsetdashtype{gp dt axes}
\gpsetlinewidth{0.50}
\draw[gp path] (0.676,1.418)--(3.721,1.418);
\gpcolor{color=gp lt color border}
\gpsetlinetype{gp lt border}
\gpsetdashtype{gp dt solid}
\gpsetlinewidth{1.00}
\draw[gp path] (0.676,1.418)--(0.766,1.418);
\node[gp node right] at (0.529,1.418) {$100$};
\gpcolor{color=gp lt color axes}
\gpsetlinetype{gp lt axes}
\gpsetdashtype{gp dt axes}
\gpsetlinewidth{0.50}
\draw[gp path] (0.676,1.925)--(3.721,1.925);
\gpcolor{color=gp lt color border}
\gpsetlinetype{gp lt border}
\gpsetdashtype{gp dt solid}
\gpsetlinewidth{1.00}
\draw[gp path] (0.676,1.925)--(0.766,1.925);
\node[gp node right] at (0.529,1.925) {$150$};
\gpcolor{color=gp lt color axes}
\gpsetlinetype{gp lt axes}
\gpsetdashtype{gp dt axes}
\gpsetlinewidth{0.50}
\draw[gp path] (0.676,2.431)--(3.721,2.431);
\gpcolor{color=gp lt color border}
\gpsetlinetype{gp lt border}
\gpsetdashtype{gp dt solid}
\gpsetlinewidth{1.00}
\draw[gp path] (0.676,2.431)--(0.766,2.431);
\node[gp node right] at (0.529,2.431) {$200$};
\gpcolor{color=gp lt color axes}
\gpsetlinetype{gp lt axes}
\gpsetdashtype{gp dt axes}
\gpsetlinewidth{0.50}
\draw[gp path] (0.676,0.405)--(0.676,2.431);
\gpcolor{color=gp lt color border}
\gpsetlinetype{gp lt border}
\gpsetdashtype{gp dt solid}
\gpsetlinewidth{1.00}
\draw[gp path] (0.676,0.405)--(0.676,0.495);
\node[gp node center] at (0.676,0.159) {$0$};
\gpcolor{color=gp lt color axes}
\gpsetlinetype{gp lt axes}
\gpsetdashtype{gp dt axes}
\gpsetlinewidth{0.50}
\draw[gp path] (1.034,0.405)--(1.034,2.431);
\gpcolor{color=gp lt color border}
\gpsetlinetype{gp lt border}
\gpsetdashtype{gp dt solid}
\gpsetlinewidth{1.00}
\draw[gp path] (1.034,0.405)--(1.034,0.495);
\node[gp node center] at (1.034,0.159) {$1$};
\gpcolor{color=gp lt color axes}
\gpsetlinetype{gp lt axes}
\gpsetdashtype{gp dt axes}
\gpsetlinewidth{0.50}
\draw[gp path] (1.392,0.405)--(1.392,2.431);
\gpcolor{color=gp lt color border}
\gpsetlinetype{gp lt border}
\gpsetdashtype{gp dt solid}
\gpsetlinewidth{1.00}
\draw[gp path] (1.392,0.405)--(1.392,0.495);
\node[gp node center] at (1.392,0.159) {$2$};
\gpcolor{color=gp lt color axes}
\gpsetlinetype{gp lt axes}
\gpsetdashtype{gp dt axes}
\gpsetlinewidth{0.50}
\draw[gp path] (1.751,0.405)--(1.751,2.431);
\gpcolor{color=gp lt color border}
\gpsetlinetype{gp lt border}
\gpsetdashtype{gp dt solid}
\gpsetlinewidth{1.00}
\draw[gp path] (1.751,0.405)--(1.751,0.495);
\node[gp node center] at (1.751,0.159) {$3$};
\gpcolor{color=gp lt color axes}
\gpsetlinetype{gp lt axes}
\gpsetdashtype{gp dt axes}
\gpsetlinewidth{0.50}
\draw[gp path] (2.109,0.405)--(2.109,2.431);
\gpcolor{color=gp lt color border}
\gpsetlinetype{gp lt border}
\gpsetdashtype{gp dt solid}
\gpsetlinewidth{1.00}
\draw[gp path] (2.109,0.405)--(2.109,0.495);
\node[gp node center] at (2.109,0.159) {$4$};
\gpcolor{color=gp lt color axes}
\gpsetlinetype{gp lt axes}
\gpsetdashtype{gp dt axes}
\gpsetlinewidth{0.50}
\draw[gp path] (2.467,0.405)--(2.467,2.431);
\gpcolor{color=gp lt color border}
\gpsetlinetype{gp lt border}
\gpsetdashtype{gp dt solid}
\gpsetlinewidth{1.00}
\draw[gp path] (2.467,0.405)--(2.467,0.495);
\node[gp node center] at (2.467,0.159) {$5$};
\gpcolor{color=gp lt color axes}
\gpsetlinetype{gp lt axes}
\gpsetdashtype{gp dt axes}
\gpsetlinewidth{0.50}
\draw[gp path] (2.825,0.405)--(2.825,2.431);
\gpcolor{color=gp lt color border}
\gpsetlinetype{gp lt border}
\gpsetdashtype{gp dt solid}
\gpsetlinewidth{1.00}
\draw[gp path] (2.825,0.405)--(2.825,0.495);
\node[gp node center] at (2.825,0.159) {$6$};
\gpcolor{color=gp lt color axes}
\gpsetlinetype{gp lt axes}
\gpsetdashtype{gp dt axes}
\gpsetlinewidth{0.50}
\draw[gp path] (3.184,0.405)--(3.184,2.431);
\gpcolor{color=gp lt color border}
\gpsetlinetype{gp lt border}
\gpsetdashtype{gp dt solid}
\gpsetlinewidth{1.00}
\draw[gp path] (3.184,0.405)--(3.184,0.495);
\node[gp node center] at (3.184,0.159) {$7$};
\gpcolor{color=gp lt color axes}
\gpsetlinetype{gp lt axes}
\gpsetdashtype{gp dt axes}
\gpsetlinewidth{0.50}
\draw[gp path] (3.542,0.405)--(3.542,2.431);
\gpcolor{color=gp lt color border}
\gpsetlinetype{gp lt border}
\gpsetdashtype{gp dt solid}
\gpsetlinewidth{1.00}
\draw[gp path] (3.542,0.405)--(3.542,0.495);
\node[gp node center] at (3.542,0.159) {$8$};
\draw[gp path] (3.721,0.405)--(3.631,0.405);
\draw[gp path] (3.721,0.912)--(3.631,0.912);
\draw[gp path] (3.721,1.418)--(3.631,1.418);
\draw[gp path] (3.721,1.925)--(3.631,1.925);
\draw[gp path] (3.721,2.431)--(3.631,2.431);
\draw[gp path] (0.676,2.431)--(0.676,0.405)--(3.721,0.405)--(3.721,2.431)--cycle;
\gpcolor{rgb color={0.392,0.561,1.000}}
\gpsetlinewidth{3.00}
\draw[gp path] (1.034,0.547)--(1.392,0.557)--(1.751,0.648)--(2.109,0.770)--(2.467,1.144)%
  --(2.825,1.185)--(3.184,1.854)--(3.542,1.965);
\gpsetpointsize{4.00}
\gp3point{gp mark 11}{}{(1.034,0.547)}
\gp3point{gp mark 11}{}{(1.392,0.557)}
\gp3point{gp mark 11}{}{(1.751,0.648)}
\gp3point{gp mark 11}{}{(2.109,0.770)}
\gp3point{gp mark 11}{}{(2.467,1.144)}
\gp3point{gp mark 11}{}{(2.825,1.185)}
\gp3point{gp mark 11}{}{(3.184,1.854)}
\gp3point{gp mark 11}{}{(3.542,1.965)}
\gpcolor{rgb color={0.863,0.149,0.498}}
\draw[gp path] (1.034,0.597)--(1.392,0.658)--(1.751,0.729)--(2.109,0.810)--(2.467,0.871)%
  --(2.825,0.871)--(3.184,0.830)--(3.542,0.841);
\gp3point{gp mark 7}{}{(1.034,0.597)}
\gp3point{gp mark 7}{}{(1.392,0.658)}
\gp3point{gp mark 7}{}{(1.751,0.729)}
\gp3point{gp mark 7}{}{(2.109,0.810)}
\gp3point{gp mark 7}{}{(2.467,0.871)}
\gp3point{gp mark 7}{}{(2.825,0.871)}
\gp3point{gp mark 7}{}{(3.184,0.830)}
\gp3point{gp mark 7}{}{(3.542,0.841)}
\gpcolor{color=gp lt color border}
\gpsetlinewidth{1.00}
\draw[gp path] (0.676,2.431)--(0.676,0.405)--(3.721,0.405)--(3.721,2.431)--cycle;
\node[gp node center,rotate=-270.0] at (-0.218,1.418) {RTT [us]};
\node[gp node center] at (2.198,-0.209) {Adv Cores};
\node[gp node center] at (2.198,2.677) {Vic 99p Lat};
\gpdefrectangularnode{gp plot 2}{\pgfpoint{0.676cm}{0.405cm}}{\pgfpoint{3.721cm}{2.431cm}}
\gpcolor{color=gp lt color axes}
\gpsetlinetype{gp lt axes}
\gpsetdashtype{gp dt axes}
\gpsetlinewidth{0.50}
\draw[gp path] (4.905,0.405)--(7.949,0.405);
\gpcolor{color=gp lt color border}
\gpsetlinetype{gp lt border}
\gpsetdashtype{gp dt solid}
\gpsetlinewidth{1.00}
\draw[gp path] (4.905,0.405)--(4.995,0.405);
\node[gp node right] at (4.758,0.405) {$0$};
\gpcolor{color=gp lt color axes}
\gpsetlinetype{gp lt axes}
\gpsetdashtype{gp dt axes}
\gpsetlinewidth{0.50}
\draw[gp path] (4.905,0.810)--(7.949,0.810);
\gpcolor{color=gp lt color border}
\gpsetlinetype{gp lt border}
\gpsetdashtype{gp dt solid}
\gpsetlinewidth{1.00}
\draw[gp path] (4.905,0.810)--(4.995,0.810);
\node[gp node right] at (4.758,0.810) {$2$};
\gpcolor{color=gp lt color axes}
\gpsetlinetype{gp lt axes}
\gpsetdashtype{gp dt axes}
\gpsetlinewidth{0.50}
\draw[gp path] (4.905,1.215)--(7.949,1.215);
\gpcolor{color=gp lt color border}
\gpsetlinetype{gp lt border}
\gpsetdashtype{gp dt solid}
\gpsetlinewidth{1.00}
\draw[gp path] (4.905,1.215)--(4.995,1.215);
\node[gp node right] at (4.758,1.215) {$4$};
\gpcolor{color=gp lt color axes}
\gpsetlinetype{gp lt axes}
\gpsetdashtype{gp dt axes}
\gpsetlinewidth{0.50}
\draw[gp path] (4.905,1.621)--(7.949,1.621);
\gpcolor{color=gp lt color border}
\gpsetlinetype{gp lt border}
\gpsetdashtype{gp dt solid}
\gpsetlinewidth{1.00}
\draw[gp path] (4.905,1.621)--(4.995,1.621);
\node[gp node right] at (4.758,1.621) {$6$};
\gpcolor{color=gp lt color axes}
\gpsetlinetype{gp lt axes}
\gpsetdashtype{gp dt axes}
\gpsetlinewidth{0.50}
\draw[gp path] (4.905,2.026)--(7.949,2.026);
\gpcolor{color=gp lt color border}
\gpsetlinetype{gp lt border}
\gpsetdashtype{gp dt solid}
\gpsetlinewidth{1.00}
\draw[gp path] (4.905,2.026)--(4.995,2.026);
\node[gp node right] at (4.758,2.026) {$8$};
\gpcolor{color=gp lt color axes}
\gpsetlinetype{gp lt axes}
\gpsetdashtype{gp dt axes}
\gpsetlinewidth{0.50}
\draw[gp path] (4.905,2.431)--(7.949,2.431);
\gpcolor{color=gp lt color border}
\gpsetlinetype{gp lt border}
\gpsetdashtype{gp dt solid}
\gpsetlinewidth{1.00}
\draw[gp path] (4.905,2.431)--(4.995,2.431);
\node[gp node right] at (4.758,2.431) {$10$};
\gpcolor{color=gp lt color axes}
\gpsetlinetype{gp lt axes}
\gpsetdashtype{gp dt axes}
\gpsetlinewidth{0.50}
\draw[gp path] (4.905,0.405)--(4.905,2.431);
\gpcolor{color=gp lt color border}
\gpsetlinetype{gp lt border}
\gpsetdashtype{gp dt solid}
\gpsetlinewidth{1.00}
\draw[gp path] (4.905,0.405)--(4.905,0.495);
\node[gp node center] at (4.905,0.159) {$0$};
\gpcolor{color=gp lt color axes}
\gpsetlinetype{gp lt axes}
\gpsetdashtype{gp dt axes}
\gpsetlinewidth{0.50}
\draw[gp path] (5.263,0.405)--(5.263,2.431);
\gpcolor{color=gp lt color border}
\gpsetlinetype{gp lt border}
\gpsetdashtype{gp dt solid}
\gpsetlinewidth{1.00}
\draw[gp path] (5.263,0.405)--(5.263,0.495);
\node[gp node center] at (5.263,0.159) {$1$};
\gpcolor{color=gp lt color axes}
\gpsetlinetype{gp lt axes}
\gpsetdashtype{gp dt axes}
\gpsetlinewidth{0.50}
\draw[gp path] (5.621,0.405)--(5.621,2.431);
\gpcolor{color=gp lt color border}
\gpsetlinetype{gp lt border}
\gpsetdashtype{gp dt solid}
\gpsetlinewidth{1.00}
\draw[gp path] (5.621,0.405)--(5.621,0.495);
\node[gp node center] at (5.621,0.159) {$2$};
\gpcolor{color=gp lt color axes}
\gpsetlinetype{gp lt axes}
\gpsetdashtype{gp dt axes}
\gpsetlinewidth{0.50}
\draw[gp path] (5.979,0.405)--(5.979,2.431);
\gpcolor{color=gp lt color border}
\gpsetlinetype{gp lt border}
\gpsetdashtype{gp dt solid}
\gpsetlinewidth{1.00}
\draw[gp path] (5.979,0.405)--(5.979,0.495);
\node[gp node center] at (5.979,0.159) {$3$};
\gpcolor{color=gp lt color axes}
\gpsetlinetype{gp lt axes}
\gpsetdashtype{gp dt axes}
\gpsetlinewidth{0.50}
\draw[gp path] (6.337,0.405)--(6.337,2.431);
\gpcolor{color=gp lt color border}
\gpsetlinetype{gp lt border}
\gpsetdashtype{gp dt solid}
\gpsetlinewidth{1.00}
\draw[gp path] (6.337,0.405)--(6.337,0.495);
\node[gp node center] at (6.337,0.159) {$4$};
\gpcolor{color=gp lt color axes}
\gpsetlinetype{gp lt axes}
\gpsetdashtype{gp dt axes}
\gpsetlinewidth{0.50}
\draw[gp path] (6.696,0.405)--(6.696,2.431);
\gpcolor{color=gp lt color border}
\gpsetlinetype{gp lt border}
\gpsetdashtype{gp dt solid}
\gpsetlinewidth{1.00}
\draw[gp path] (6.696,0.405)--(6.696,0.495);
\node[gp node center] at (6.696,0.159) {$5$};
\gpcolor{color=gp lt color axes}
\gpsetlinetype{gp lt axes}
\gpsetdashtype{gp dt axes}
\gpsetlinewidth{0.50}
\draw[gp path] (7.054,0.405)--(7.054,2.431);
\gpcolor{color=gp lt color border}
\gpsetlinetype{gp lt border}
\gpsetdashtype{gp dt solid}
\gpsetlinewidth{1.00}
\draw[gp path] (7.054,0.405)--(7.054,0.495);
\node[gp node center] at (7.054,0.159) {$6$};
\gpcolor{color=gp lt color axes}
\gpsetlinetype{gp lt axes}
\gpsetdashtype{gp dt axes}
\gpsetlinewidth{0.50}
\draw[gp path] (7.412,0.405)--(7.412,2.431);
\gpcolor{color=gp lt color border}
\gpsetlinetype{gp lt border}
\gpsetdashtype{gp dt solid}
\gpsetlinewidth{1.00}
\draw[gp path] (7.412,0.405)--(7.412,0.495);
\node[gp node center] at (7.412,0.159) {$7$};
\gpcolor{color=gp lt color axes}
\gpsetlinetype{gp lt axes}
\gpsetdashtype{gp dt axes}
\gpsetlinewidth{0.50}
\draw[gp path] (7.770,0.405)--(7.770,2.431);
\gpcolor{color=gp lt color border}
\gpsetlinetype{gp lt border}
\gpsetdashtype{gp dt solid}
\gpsetlinewidth{1.00}
\draw[gp path] (7.770,0.405)--(7.770,0.495);
\node[gp node center] at (7.770,0.159) {$8$};
\draw[gp path] (7.949,0.405)--(7.859,0.405);
\draw[gp path] (7.949,0.810)--(7.859,0.810);
\draw[gp path] (7.949,1.215)--(7.859,1.215);
\draw[gp path] (7.949,1.621)--(7.859,1.621);
\draw[gp path] (7.949,2.026)--(7.859,2.026);
\draw[gp path] (7.949,2.431)--(7.859,2.431);
\draw[gp path] (4.905,2.431)--(4.905,0.405)--(7.949,0.405)--(7.949,2.431)--cycle;
\gpcolor{rgb color={0.863,0.149,0.498}}
\gpsetlinewidth{3.00}
\draw[gp path] (5.263,0.603)--(5.621,0.603)--(5.979,0.603)--(6.337,0.603)--(6.696,0.603)%
  --(7.054,0.603)--(7.412,0.603)--(7.770,0.603);
\gp3point{gp mark 7}{}{(5.263,0.603)}
\gp3point{gp mark 7}{}{(5.621,0.603)}
\gp3point{gp mark 7}{}{(5.979,0.603)}
\gp3point{gp mark 7}{}{(6.337,0.603)}
\gp3point{gp mark 7}{}{(6.696,0.603)}
\gp3point{gp mark 7}{}{(7.054,0.603)}
\gp3point{gp mark 7}{}{(7.412,0.603)}
\gp3point{gp mark 7}{}{(7.770,0.603)}
\gpcolor{rgb color={0.392,0.561,1.000}}
\draw[gp path] (5.263,0.603)--(5.621,0.603)--(5.979,0.603)--(6.337,0.603)--(6.696,0.603)%
  --(7.054,0.603)--(7.412,0.603)--(7.770,0.602);
\gp3point{gp mark 11}{}{(5.263,0.603)}
\gp3point{gp mark 11}{}{(5.621,0.603)}
\gp3point{gp mark 11}{}{(5.979,0.603)}
\gp3point{gp mark 11}{}{(6.337,0.603)}
\gp3point{gp mark 11}{}{(6.696,0.603)}
\gp3point{gp mark 11}{}{(7.054,0.603)}
\gp3point{gp mark 11}{}{(7.412,0.603)}
\gp3point{gp mark 11}{}{(7.770,0.602)}
\gpcolor{rgb color={0.471,0.369,0.941}}
\draw[gp path] (5.263,1.196)--(5.621,1.459)--(5.979,1.765)--(6.337,1.876)--(6.696,2.141)%
  --(7.054,2.108)--(7.412,2.127)--(7.770,2.157);
\gp3point{gp mark 5}{}{(5.263,1.196)}
\gp3point{gp mark 5}{}{(5.621,1.459)}
\gp3point{gp mark 5}{}{(5.979,1.765)}
\gp3point{gp mark 5}{}{(6.337,1.876)}
\gp3point{gp mark 5}{}{(6.696,2.141)}
\gp3point{gp mark 5}{}{(7.054,2.108)}
\gp3point{gp mark 5}{}{(7.412,2.127)}
\gp3point{gp mark 5}{}{(7.770,2.157)}
\gpcolor{rgb color={0.996,0.380,0.000}}
\draw[gp path] (5.263,1.196)--(5.621,1.456)--(5.979,1.762)--(6.337,1.802)--(6.696,1.814)%
  --(7.054,1.817)--(7.412,1.816)--(7.770,1.840);
\gp3point{gp mark 9}{}{(5.263,1.196)}
\gp3point{gp mark 9}{}{(5.621,1.456)}
\gp3point{gp mark 9}{}{(5.979,1.762)}
\gp3point{gp mark 9}{}{(6.337,1.802)}
\gp3point{gp mark 9}{}{(6.696,1.814)}
\gp3point{gp mark 9}{}{(7.054,1.817)}
\gp3point{gp mark 9}{}{(7.412,1.816)}
\gp3point{gp mark 9}{}{(7.770,1.840)}
\gpcolor{color=gp lt color border}
\gpsetlinewidth{1.00}
\draw[gp path] (4.905,2.431)--(4.905,0.405)--(7.949,0.405)--(7.949,2.431)--cycle;
\node[gp node center,rotate=-270.0] at (4.157,1.418) {MReq/s};
\node[gp node center] at (6.427,-0.209) {Adv Cores};
\node[gp node center] at (6.427,2.677) {Vic and Adv Tput};
\gpdefrectangularnode{gp plot 3}{\pgfpoint{4.905cm}{0.405cm}}{\pgfpoint{7.949cm}{2.431cm}}
\end{tikzpicture}
  \caption{A full TCP adversary inflates victim 99p RTT to 154\,\textmu s
  without isolation, but only 46\,\textmu s with isolation.}
\label{fig:perf_iso}%
\end{figure}

As the adversary increases its core count, \autoref{fig:perf_iso}
shows that disabling isolation
causes severe degradation in the victim's 99th-percentile RTT, which
rises from 14\,\textmu s to 154\,\textmu s. Enabling isolation bounds it
at 46\,\textmu s.
The system transitions from unloaded to saturated between
one and three cores. In this range, the victim's 99th-percentile RTT increases even
with isolation, as the adversary scales its throughput to reach
its fair share of resources; it
rises from 19\,\textmu s to 40\,\textmu s despite isolation, but remains
bounded as the adversary attempts to generate interference
with more than three cores.

\subsection{Static Bounds Are Necessary, but Fit Complex Protocols}
\label{ssec:eval:static_bounds}

Runtime accounting alone does not determine how much interference a tenant can
impose: larger handlers still consume longer bursts of fast path time before
the non-preemptive scheduler can react. To quantify this effect, we deploy the
baseline UDP implementation as the victim. The adversary is a
synthetic variant of the same UDP protocol whose |event_deq| handler we pad
to different total eBPF instruction counts. We measure the victim's
99th-percentile RTT and throughput as the adversary's handler size grows.

For reference, TCP uses 690 instructions for |event_sched|, 1768 for |event_rx|,
and 339 for |event_deq|. UDP uses 334 instructions for |event_rx| and 698 for
|event_deq|. UDP does not use |event_sched|, as it emits packets directly in
|event_deq| instead of staging them in the scheduler.

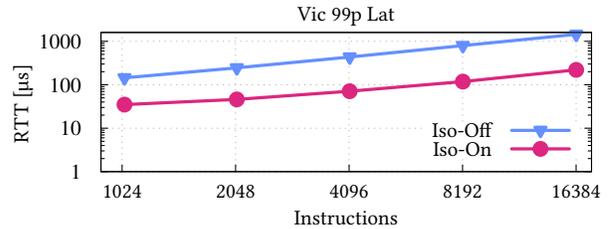
\begin{figure}%
\centering%
\begin{tikzpicture}[gnuplot]
\tikzset{every node/.append style={font={\fontsize{8.0pt}{9.6pt}\selectfont}}}
\path (0.000,0.000) rectangle (8.458,3.378);
\gpcolor{color=gp lt color axes}
\gpsetlinetype{gp lt axes}
\gpsetdashtype{gp dt axes}
\gpsetlinewidth{0.50}
\draw[gp path] (1.201,0.787)--(7.722,0.787);
\gpcolor{color=gp lt color border}
\gpsetlinetype{gp lt border}
\gpsetdashtype{gp dt solid}
\gpsetlinewidth{1.00}
\draw[gp path] (1.201,0.787)--(1.291,0.787);
\node[gp node right] at (1.054,0.787) {$1$};
\draw[gp path] (1.201,0.961)--(1.246,0.961);
\draw[gp path] (1.201,1.063)--(1.246,1.063);
\draw[gp path] (1.201,1.135)--(1.246,1.135);
\draw[gp path] (1.201,1.191)--(1.246,1.191);
\draw[gp path] (1.201,1.237)--(1.246,1.237);
\draw[gp path] (1.201,1.275)--(1.246,1.275);
\draw[gp path] (1.201,1.309)--(1.246,1.309);
\draw[gp path] (1.201,1.339)--(1.246,1.339);
\gpcolor{color=gp lt color axes}
\gpsetlinetype{gp lt axes}
\gpsetdashtype{gp dt axes}
\gpsetlinewidth{0.50}
\draw[gp path] (1.201,1.365)--(5.484,1.365);
\draw[gp path] (7.575,1.365)--(7.722,1.365);
\gpcolor{color=gp lt color border}
\gpsetlinetype{gp lt border}
\gpsetdashtype{gp dt solid}
\gpsetlinewidth{1.00}
\draw[gp path] (1.201,1.365)--(1.291,1.365);
\node[gp node right] at (1.054,1.365) {$10$};
\draw[gp path] (1.201,1.539)--(1.246,1.539);
\draw[gp path] (1.201,1.641)--(1.246,1.641);
\draw[gp path] (1.201,1.713)--(1.246,1.713);
\draw[gp path] (1.201,1.769)--(1.246,1.769);
\draw[gp path] (1.201,1.815)--(1.246,1.815);
\draw[gp path] (1.201,1.853)--(1.246,1.853);
\draw[gp path] (1.201,1.887)--(1.246,1.887);
\draw[gp path] (1.201,1.917)--(1.246,1.917);
\gpcolor{color=gp lt color axes}
\gpsetlinetype{gp lt axes}
\gpsetdashtype{gp dt axes}
\gpsetlinewidth{0.50}
\draw[gp path] (1.201,1.943)--(7.722,1.943);
\gpcolor{color=gp lt color border}
\gpsetlinetype{gp lt border}
\gpsetdashtype{gp dt solid}
\gpsetlinewidth{1.00}
\draw[gp path] (1.201,1.943)--(1.291,1.943);
\node[gp node right] at (1.054,1.943) {$100$};
\draw[gp path] (1.201,2.117)--(1.246,2.117);
\draw[gp path] (1.201,2.219)--(1.246,2.219);
\draw[gp path] (1.201,2.291)--(1.246,2.291);
\draw[gp path] (1.201,2.347)--(1.246,2.347);
\draw[gp path] (1.201,2.393)--(1.246,2.393);
\draw[gp path] (1.201,2.431)--(1.246,2.431);
\draw[gp path] (1.201,2.465)--(1.246,2.465);
\draw[gp path] (1.201,2.495)--(1.246,2.495);
\gpcolor{color=gp lt color axes}
\gpsetlinetype{gp lt axes}
\gpsetdashtype{gp dt axes}
\gpsetlinewidth{0.50}
\draw[gp path] (1.201,2.521)--(7.722,2.521);
\gpcolor{color=gp lt color border}
\gpsetlinetype{gp lt border}
\gpsetdashtype{gp dt solid}
\gpsetlinewidth{1.00}
\draw[gp path] (1.201,2.521)--(1.291,2.521);
\node[gp node right] at (1.054,2.521) {$1000$};
\gpcolor{color=gp lt color axes}
\gpsetlinetype{gp lt axes}
\gpsetdashtype{gp dt axes}
\gpsetlinewidth{0.50}
\draw[gp path] (1.482,0.787)--(1.482,2.639);
\gpcolor{color=gp lt color border}
\gpsetlinetype{gp lt border}
\gpsetdashtype{gp dt solid}
\gpsetlinewidth{1.00}
\draw[gp path] (1.482,0.787)--(1.482,0.877);
\node[gp node center] at (1.482,0.541) {1024};
\gpcolor{color=gp lt color axes}
\gpsetlinetype{gp lt axes}
\gpsetdashtype{gp dt axes}
\gpsetlinewidth{0.50}
\draw[gp path] (2.991,0.787)--(2.991,2.639);
\gpcolor{color=gp lt color border}
\gpsetlinetype{gp lt border}
\gpsetdashtype{gp dt solid}
\gpsetlinewidth{1.00}
\draw[gp path] (2.991,0.787)--(2.991,0.877);
\node[gp node center] at (2.991,0.541) {2048};
\gpcolor{color=gp lt color axes}
\gpsetlinetype{gp lt axes}
\gpsetdashtype{gp dt axes}
\gpsetlinewidth{0.50}
\draw[gp path] (4.500,0.787)--(4.500,2.639);
\gpcolor{color=gp lt color border}
\gpsetlinetype{gp lt border}
\gpsetdashtype{gp dt solid}
\gpsetlinewidth{1.00}
\draw[gp path] (4.500,0.787)--(4.500,0.877);
\node[gp node center] at (4.500,0.541) {4096};
\gpcolor{color=gp lt color axes}
\gpsetlinetype{gp lt axes}
\gpsetdashtype{gp dt axes}
\gpsetlinewidth{0.50}
\draw[gp path] (6.008,0.787)--(6.008,0.967);
\draw[gp path] (6.008,1.459)--(6.008,2.639);
\gpcolor{color=gp lt color border}
\gpsetlinetype{gp lt border}
\gpsetdashtype{gp dt solid}
\gpsetlinewidth{1.00}
\draw[gp path] (6.008,0.787)--(6.008,0.877);
\node[gp node center] at (6.008,0.541) {8192};
\gpcolor{color=gp lt color axes}
\gpsetlinetype{gp lt axes}
\gpsetdashtype{gp dt axes}
\gpsetlinewidth{0.50}
\draw[gp path] (7.517,0.787)--(7.517,0.967);
\draw[gp path] (7.517,1.459)--(7.517,2.639);
\gpcolor{color=gp lt color border}
\gpsetlinetype{gp lt border}
\gpsetdashtype{gp dt solid}
\gpsetlinewidth{1.00}
\draw[gp path] (7.517,0.787)--(7.517,0.877);
\node[gp node center] at (7.517,0.541) {16384};
\draw[gp path] (7.722,0.787)--(7.632,0.787);
\draw[gp path] (7.722,0.961)--(7.677,0.961);
\draw[gp path] (7.722,1.063)--(7.677,1.063);
\draw[gp path] (7.722,1.135)--(7.677,1.135);
\draw[gp path] (7.722,1.191)--(7.677,1.191);
\draw[gp path] (7.722,1.237)--(7.677,1.237);
\draw[gp path] (7.722,1.275)--(7.677,1.275);
\draw[gp path] (7.722,1.309)--(7.677,1.309);
\draw[gp path] (7.722,1.339)--(7.677,1.339);
\draw[gp path] (7.722,1.365)--(7.632,1.365);
\draw[gp path] (7.722,1.539)--(7.677,1.539);
\draw[gp path] (7.722,1.641)--(7.677,1.641);
\draw[gp path] (7.722,1.713)--(7.677,1.713);
\draw[gp path] (7.722,1.769)--(7.677,1.769);
\draw[gp path] (7.722,1.815)--(7.677,1.815);
\draw[gp path] (7.722,1.853)--(7.677,1.853);
\draw[gp path] (7.722,1.887)--(7.677,1.887);
\draw[gp path] (7.722,1.917)--(7.677,1.917);
\draw[gp path] (7.722,1.943)--(7.632,1.943);
\draw[gp path] (7.722,2.117)--(7.677,2.117);
\draw[gp path] (7.722,2.219)--(7.677,2.219);
\draw[gp path] (7.722,2.291)--(7.677,2.291);
\draw[gp path] (7.722,2.347)--(7.677,2.347);
\draw[gp path] (7.722,2.393)--(7.677,2.393);
\draw[gp path] (7.722,2.431)--(7.677,2.431);
\draw[gp path] (7.722,2.465)--(7.677,2.465);
\draw[gp path] (7.722,2.495)--(7.677,2.495);
\draw[gp path] (7.722,2.521)--(7.632,2.521);
\draw[gp path] (1.201,2.639)--(1.201,0.787)--(7.722,0.787)--(7.722,2.639)--cycle;
\node[gp node right] at (6.513,1.336) {Iso-Off};
\gpcolor{rgb color={0.392,0.561,1.000}}
\gpsetlinewidth{3.00}
\draw[gp path] (6.660,1.336)--(7.428,1.336);
\draw[gp path] (1.514,2.035)--(3.007,2.168)--(4.508,2.312)--(6.012,2.464)--(7.519,2.614);
\gpsetpointsize{6.00}
\gp3point{gp mark 11}{}{(1.514,2.035)}
\gp3point{gp mark 11}{}{(3.007,2.168)}
\gp3point{gp mark 11}{}{(4.508,2.312)}
\gp3point{gp mark 11}{}{(6.012,2.464)}
\gp3point{gp mark 11}{}{(7.519,2.614)}
\gp3point{gp mark 11}{}{(7.044,1.336)}
\gpcolor{color=gp lt color border}
\node[gp node right] at (6.513,1.090) {Iso-On};
\gpcolor{rgb color={0.863,0.149,0.498}}
\draw[gp path] (6.660,1.090)--(7.428,1.090);
\draw[gp path] (1.514,1.679)--(3.007,1.748)--(4.508,1.857)--(6.012,1.985)--(7.519,2.141);
\gp3point{gp mark 7}{}{(1.514,1.679)}
\gp3point{gp mark 7}{}{(3.007,1.748)}
\gp3point{gp mark 7}{}{(4.508,1.857)}
\gp3point{gp mark 7}{}{(6.012,1.985)}
\gp3point{gp mark 7}{}{(7.519,2.141)}
\gp3point{gp mark 7}{}{(7.044,1.090)}
\gpcolor{color=gp lt color border}
\gpsetlinewidth{1.00}
\draw[gp path] (1.201,2.639)--(1.201,0.787)--(7.722,0.787)--(7.722,2.639)--cycle;
\node[gp node center,rotate=-270.0] at (0.159,1.713) {RTT [µs]};
\node[gp node center] at (4.461,0.172) {Instructions};
\node[gp node center] at (4.461,2.885) {Vic 99p Lat};
\gpdefrectangularnode{gp plot 1}{\pgfpoint{1.201cm}{0.787cm}}{\pgfpoint{7.722cm}{2.639cm}}
\end{tikzpicture}
  \caption{Interference grows with handler size even with isolation,
  motivating static bounds; TCP fits below 2,048.}
\label{fig:iso_sens}%
\end{figure}

\autoref{fig:iso_sens} shows that increasing the adversary's program
size increases the victim's 99th-percentile RTT even with cycle-level
isolation. It rises from 35\,\textmu s at 1,024 instructions to
220\,\textmu s at 16,384 instructions, 
corresponding to a 6.29$\times$ increase.
Without isolation, the degradation is significantly larger: the victim's
99th-percentile RTT rises from 144\,\textmu s to 1449\,\textmu s over the
same range.
These results show that runtime accounting alone is not sufficient:
admission-time instruction bounds are necessary to complement cycle-level
isolation. At the same time, the required bounds remain permissive enough for
realistic full protocols. In this experiment, programs of up to 2,048
instructions incur only minor residual interference, and \sys's largest TCP
handler remains below that regime at 1,768 instructions.

\section{Related Work}%
\label{sec:related}

\paragraph{Shared multi-tenant network stacks.}
Virtuoso~\cite{stolet:virtuoso}, Junction~\cite{fried:junction},
Andromeda~\cite{dalton:andromeda}, Snap~\cite{marty:snap}, and
NetKernel~\cite{niu:netkernel},  
centralize the data path and multiplex many tenants, while
enforcing varying levels of performance isolation in a shared stack.
However, they treat the shared stack implementation
as operator-controlled: tenants cannot inject their own protocol
logic and must use the provider's protocol implementation.
\sys targets the missing design point these systems leave open: it preserves the
efficiency and isolation benefits of a shared host stack while allowing
tenants to deploy their own bounded protocol handlers.

\paragraph{eBPF-based extensible packet processing.}
Systems such as XDP~\cite{jorgensen:xdp},
and Flexnet~\cite{dwivedi:flexnet} extend the monolithic kernel
stack and allow developers to customize packet processing and
implement high-performance functions inside the kernel.
Nonetheless, they expose hook-sized interfaces rather than a bounded contract
for full protocols, and they do not address tenant isolation in a shared
multi-tenant host datapath.
eTran~\cite{chen:etran} extends full transport protocols
by using eBPF and adding new hook points to the kernel. However, it does not
address performance isolation in the multi-tenant shared-stack setting and
still treats tenant eBPF as an opaque JIT-compiled extension boundary.
MorphOS~\cite{okelmann:morphos} brings verified eBPF extensibility 
to high-performance unikernel-based OSes, but it limits itself
to customizing existing applications, such as
Click~\cite{kohler:click} VNF, and its interface is not
designed for full tenant-defined protocols in a shared cloud datapath.

\paragraph{eBPF in virtualised environments.}
Prior work has used eBPF in different ways to accelerate virtualised
environments.
Hyperupcalls~\cite{amit:hyperupcalls} provide a framework
to offload guest functionality to a hypervisor via eBPF
and HyperTurtle~\cite{zur:hyperturtle} leverages them
to accelerate nested virtualisation by safely offloading
parts of the \texttt{vm-exit} logic from the nested VM to
the virtualised or bare-metal hypervisor.
Additionally, RosenBridge~\cite{qiu:rosenbridge} focuses on accelerating
the storage I/O datapath across virtualisation boundaries. 
It uses eBPF to create an execution runtime in the host where
guests can offload near-data processing optimisations
to the bare-metal I/O datapath.
Likewise, EXO~\cite{wang:exo} enhances the performance of 
storage paravirtualisation by extending the kernel to query
guest-to-host address mappings without switching
to QEMU in userspace for backend processing.
These systems largely use eBPF to accelerate specific
virtualization mechanisms. In contrast, \sys exposes
a bounded interface that makes the
host datapath safely programmable for untrusted
tenants in a virtualised environment, allowing full
tenant-defined protocols to execute on a shared network stack.

\paragraph{eBPF performance optimisations.}
Morpheus~\cite{miano:morpheus} improves the performance
of packet processing programs that use eBPF by collecting
runtime data to specialise compiler optimisations applied by
the JIT. Similarly, Merlin~\cite{mao:merlin} adds eBPF-aware LLVM
passes to the build process to emit more optimised code and
K2~\cite{xu:k2} uses stochastic search to find a more efficient
eBPF program. These approaches improve the quality of the
generated programs, but still treat them
as opaque and forgo optimisation opportunities between
the host datapath and tenant eBPF handlers.
KFuse~\cite{kuo:kfuse} takes a broader approach and merges
chains of eBPF programs to jointly optimise the generated
code for all programs in the chain. Nonetheless, it still
misses optimisation opportunities across both the host-program boundary
and co-resident tenants in a shared fast path.

\paragraph{Cross-module compiler optimisations.}
Compiler frameworks such as ThinLTO~\cite{johnson:thinlto}, 
and BOLT~\cite{panchenko:bolt} 
combine modules or compilation units into a single optimisation scope.
These approaches show the value of enlarging optimisation scope,
but they target static source modules.
\sys applies the idea online to a multi-tenant network stack
and recompiles a unified LLVM module as new guests join and 
leave the system, enabling optimisation across tenant/provider and
tenant/tenant boundaries in the shared fast path.

\paragraph{Performance isolation.}
Prior work tackles performance interference across tenants
using different mechanisms. KFlex~\cite{dwivedi:kflex}
handles noisy eBPF neighbors by terminating long-running
programs once execution exceeds a fixed quantum, treating
isolation as a coarse-grained, reactive mechanism. \sys,
in contrast, enforces fine-grained isolation via per-tenant
cycle-level accounting and scheduling across bounded
datapath phases, reducing disruption and overhead.
Virtuoso uses fine-grained datapath accounting and scheduling to mitigate
interference between tenants in a fixed shared stack. \sys builds on this
mechanism, but tackles the additional problem that tenant-defined handlers
execute in the same datapath and have much more variable cost. The key
distinction is that Virtuoso isolates a fixed shared stack, whereas \sys
preserves isolation even when the shared fast path executes tenant-defined
handlers for full protocols.

\section{Conclusion}
\sys shows that shared cloud network stacks need not choose between
shared-stack efficiency and tenant-defined full protocols. It exposes a
bounded host-side fast path---receive, scheduler-triggered send, and dequeue
handlers---paired with a tenant slow path, allowing tenants to implement full
protocols while keeping the trusted host interface narrow. Joint compilation across tenant
handlers, provider infrastructure, and co-resident tenants recovers most of
the programmability tax of eBPF extensions, while runtime cycle accounting
preserves strong isolation even when the shared fast path executes
tenant-defined handlers. Together, these results show that provider-managed
shared stacks can remain efficient, predictable, and programmable in
multi-tenant clouds.
 \if \ANON 0
\section*{Acknowledgements}
We thank Srushti Singh for her contributions to
an early implementation of the eBPF bytecode
upload.

 \fi

\bibliographystyle{plain}
\bibliography{paper,bibdb/strings,bibdb/papers,bibdb/defs}

\label{page:last}
\end{document}